\documentclass[ALICE,manyauthors]{cernphprep}

\usepackage[comma,square,numbers,sort&compress]{natbib}
\usepackage{hyperref}
\usepackage{lineno}
\usepackage[T1]{fontenc}

\usepackage{grffile}
\usepackage{rotating}
\usepackage{subfigure}
\usepackage{multirow}
\usepackage{multicol}
\usepackage{relsize} 
\usepackage{float} 
\usepackage{verbatim} 
\usepackage[section]{placeins} 
\usepackage{enumitem}
\usepackage{blindtext} 
\usepackage{color, soul}
\usepackage{pdflscape}

\newcommand{\degrees}{^{\rm o}}

\newcommand{\pt}{\ensuremath{p_{\rm T}}\xspace}
\newcommand{\ptabove}[1]{\ensuremath{p_{\rm T}~>~#1}~GeV/$c$}

\newcommand{\ptjetrange}[2]{\ensuremath{#1~<~p_{\rm T}^{\rm jet}~<~#2}~GeV/$c$}

\newcommand{\akT}{anti-$k_T$\xspace}

\newcommand{\pp}{pp\xspace}
\newcommand{\PbPb}{Pb--Pb\xspace}
\newcommand{\AuAu}{Au--Au\xspace}

\newcommand{\sref}[1]{Sec.~\ref{#1}}
\newcommand{\fref}[1]{Fig.~\ref{#1}}
\newcommand{\tref}[1]{Tab.~\ref{#1}}
\newcommand{\eref}[1]{equation~\ref{#1}}

\newcommand{\Fref}[1]{Figure~\ref{#1}}
\newcommand{\Tref}[1]{Table~\ref{#1}}
\newcommand{\Eref}[1]{Equation~\ref{#1}}

\newcommand{\dphi}{$\Delta\phi$\xspace}
\newcommand{\deta}{$\Delta\eta$\xspace}
\newcommand{\GeV}{GeV/$c$\xspace}

\newcommand{\sNN}{$\sqrt{s_{\mathrm{NN}}}$\xspace}
\newcommand{\ns}{near-side\xspace}
\newcommand{\as}{away-side\xspace}
\newcommand{\ptassoc}{$p_{\mathrm T}^{\mathrm{assoc}}$\xspace}
\newcommand{\pttrigrange}[2]{\ensuremath{#1~<~p_{\rm T}^{\mathrm{jet}}~<~#2}~GeV/$c$}
\newcommand{\ptassocrange}[2]{~\ensuremath{#1~<~p_{\rm T}^{\mathrm{assoc}}~<~#2}~GeV/$c$}

\newcommand{\vn}{$v_{n}$\xspace}

\newcommand{\vnum}[1]{$v_{#1}$\xspace}

\newcommand\Tstrut{\rule{0pt}{2.6ex}}         
\newcommand\Bstrut{\rule[-0.9ex]{0pt}{0pt}}

\begin{document}
\begin{titlepage}
\PHyear{2019}
\PHnumber{244}      
\PHdate{28 October}  

\title{Jet-hadron correlations measured relative to the second order event plane in Pb--Pb collisions at \sNN = 2.76 TeV}
\ShortTitle{Jet-hadron correlations relative to the event plane}   

\Collaboration{ALICE Collaboration\thanks{See Appendix~\ref{app:collab} for the list of collaboration members}}
\ShortAuthor{ALICE Collaboration}

\begin{abstract}
The Quark Gluon Plasma (QGP) produced in ultra relativistic heavy-ion collisions  at the Large Hadron Collider (LHC) can be studied by measuring the modifications of jets formed by hard scattered partons which interact with the medium.
We studied these modifications via angular correlations of jets with charged hadrons for jets with momenta \pttrigrange{20}{40} as a function of the associated particle momentum.  The reaction plane fit (RPF) method is used in this analysis to remove the flow modulated background.  The analysis of angular correlations for different orientations of the jet relative to the second order event plane allows for the study of the path length dependence of medium modifications to jets.  We present the dependence of azimuthal angular correlations of charged hadrons with respect to the angle of the axis of a reconstructed jet relative to the event plane in \PbPb collisions at \sNN = 2.76 TeV.  
The dependence of particle yields associated with jets on the angle of the jet with respect to the event plane is presented.  Correlations at different angles relative to the event plane are compared through ratios and differences of the yield.  No dependence of the results on the angle of the jet with respect to the event plane is observed within uncertainties, which is consistent with no significant path length dependence of the medium modifications for this observable.

\end{abstract}
\end{titlepage}
\setcounter{page}{2}

\section{Introduction}\label{Sec:Introduction}
A hot, dense liquid of quarks and gluons is created in high energy collisions of heavy ions at the Relativistic Heavy Ion Collider (RHIC)~\cite{Adcox:2004mh,Adams:2005dq,Back:2004je,Arsene:2004fa} and the Large Hadron Collider (LHC)~\cite{Aamodt:2011mr,Aamodt:2010cz,Aamodt:2010jd,Aamodt:2010pb,Aamodt:2010pa,Aad:2010bu,Chatrchyan:2011sx}.  This strongly interacting medium, called the Quark Gluon Plasma (QGP), suppresses colored probes such as quarks and gluons.

Hard parton scatterings occur early in the collision and lead to the production of jets, collimated sprays of particles formed from the fragmentation of the scattered parton.  These hard partons lose energy through induced gluon bremsstrahlung and elastic collisions with medium partons as they traverse the QGP, leading to a broadening of the resulting jet and softening of its constituents~\cite{Connors:2017ptx,Qin:2015srf}. This energy loss can be studied with measurements of high transverse momentum hadrons or reconstructed jets. 
High momentum charged hadron production is suppressed by a factor of approximately five in \AuAu collisions at RHIC~\cite{Adams:2003kv,Adler:2003qi,Back:2004bq} and up to a factor of nearly ten in \PbPb collisions at the LHC~\cite{Aamodt:2010jd,CMS:2012aa} relative to that in \pp collisions. These measurements are used to constrain the transport coefficient $\hat{q}$, the squared partonic momentum exchanged with the medium divided by the path length traversed~\cite{Burke:2013yra} in the QGP.  

An enhancement of particle production at low \pt due to medium interactions has been observed with measurements of fragmentation functions, the momentum distributions of particles within the jet~\cite{Aad:2014wha,Chatrchyan:2014ava,Chatrchyan:2012gw},  as well as through broadening in high momentum dihadron correlations~\cite{Adare:2012qi,Agakishiev:2011st} and jet-hadron correlations~\cite{Adamczyk:2013jei,Khachatryan:2016erx,Chatrchyan:2011sx,Acharya:2019ssy}.

Measurements of correlations allow studies of lower energy jets and of the soft constituents by means of statistical subtraction of the large combinatorial background at lower momenta.  Studies of correlations have been limited by methods for background subtraction due to the structures in the background correlated with the signal because of hydrodynamical flow.  The recent development of the Reaction Plane Fit (RPF) method enables precision subtraction of the background for both jet-hadron and dihadron correlations~\cite{Sharma:2015qra}.

The path length dependence of partonic energy loss can be constrained through measurements of the dependence of azimuthal correlations of high momentum particles or reconstructed jets on the angle of the jet relative to second order event plane of the collision.  Because the overlap region of the incoming nuclei for non-central collisions is almond shaped, particles traveling perpendicular to this event plane (out-of-plane) have a longer path length through the medium on average than those traveling in the direction of the event plane (in-plane). Therefore, the suppression of high momentum single particles is expected to be greater in the out-of-plane direction than in-plane~\cite{Afanasiev:2009aa}. This is also evident in the azimuthal anisotropy of high-$p_{\rm T}$ single particle~\cite{Adare:2013wop,Abelev:2012di,Chatrchyan:2012xq,Acharya:2018zuq} 
and jet~\cite{Adam:2015mda,Aad:2013sla} production relative to the second order event plane.  This suppression indicates that there are fewer jets out-of-plane after interactions with the medium, but is not a measure of the properties of the surviving jets relative to that plane.  
Measurements of dihadron correlations relative to the event plane at RHIC indicate suppression and some broadening~\cite{Adare:2010mq,Agakishiev:2014ada}, but do not exhibit much event plane dependence~\cite{Nattrass:2016cln}.  Some theoretical studies indicate that jet-by-jet fluctuations in the energy loss may be as important as the path length dependence for some observables, such as azimuthal anisotropies at high momentum and di-jet asymmetries~\cite{Zapp:2013zya,Noronha-Hostler:2016eow,Betz:2016ayq}.

Measurements for the event plane dependence of jet modification can therefore provide insight into the relative importance of path length for partonic energy loss. We present  measurements of jet-hadron correlations relative to the event plane in \PbPb collisions at \sNN = 2.76 TeV using A Large Ion Collider Experiment (ALICE) detector.  We first describe the details of the measurement technique and then present the results.  We conclude with a discussion of the constraints these measurements provide for models.
 
\section{The ALICE detector}\label{Sec:Detector}

A detailed description of the ALICE detector and its subdetectors can be found in~\cite{Abelev:2014ffa}. The detectors used for the present analysis are briefly described in this section. These are the forward scintillator arrays (V0)~\cite{Abbas:2013taa,Cortese:2004aa}, Inner Tracking System (ITS)~\cite{Aamodt:2010aa}, the Time Projection Chamber (TPC)~\cite{Alme:2010ke}, and the Electromagnetic Calorimeter (EMCal)~\cite{Abeysekara:2010ze}.  

The V0 detector is used for centrality estimation and event plane reconstruction. The V0 system consists of two scintillator arrays located at asymmetric positions, one at a pseudorapidity range of $2.8<\eta<5.1$ (V0A) and the other at $-3.7<\eta<-1.7$ (V0C)~\cite{Abbas:2013taa}.  Each set of arrays is made of four radial rings with each ring divided into eight sections in the azimuthal direction~\cite{Abbas:2013taa}.

The ITS and TPC detectors provide tracking of charged particles over the full range of azimuth with a pseudorapidity range of $|\eta| < 0.9$.  They are located inside the central barrel solenoidal magnet which provides a homogenous field with strength of 0.5 T~\cite{Adam:2015ewa}.  The ITS is a cylindrical silicon detector made up of 6 layers located at the center of the main-barrel.  The first two layers are the Silicon Pixel Detectors (SPD), followed by two layers of Silicon Drift Detectors (SDD), and two outer layers of Silicon Strip Detectors (SSD)~\cite{Dellacasa:1999kf}. The TPC surrounds the ITS and is the main detector used for tracking in ALICE. It is filled with a gas mixture of Ne and CO$_{2}$~\cite{Dellacasa:2000bm}. The transverse momentum and charge of the particles can be inferred from the curvature of the tracks. Combining information from the ITS and TPC allows for the momentum determination of charged particles as low as \pt~$\approx$~0.15~\GeV up to \pt~$\approx$~100~\GeV.

Track selection is optimized for track quality, momentum resolution, and nearly uniform azimuthal acceptance, as in~\cite{Adam:2015ewa}.  At least three hits in the ITS are required. Tracks without a hit in the SPD are refit to include the primary vertex, reducing the azimuthal dependence of the track reconstruction efficiency while maintaining good momentum resolution.  The tracks used in this analysis are required to have 80\% of the geometrically allowed space points and at least 70 total space points in the TPC.  The tracking efficiency is determined from simulations of the detector response using tracks simulated with {\sc hijing}~\cite{Wang:1991hta} propagated through the detector using {\sc geant3}~\cite{Brun:1994aa} and ranges from 80--85\% in the momentum range used in this analysis.

The uncertainty on the single track reconstruction efficiency is 4\%, with an additional 1\% systematic uncertainty due to contamination from secondary tracks~\cite{Adam:2016xbp,Adam:2015ewa,Abelev:2013fn,Abelev:2013kqa}.  This uncertainty is correlated point-to-point and contributes to the scale uncertainty in the correlation functions and yields.

The EMCal is used for the neutral energy reconstruction and triggering. It is a lead-scintillator sampling calorimeter with a pseudorapidity coverage of $|\eta| < 0.7$ and an active azimuthal range of $\Delta \phi = 107\degrees$ in the readout in the 2011 configuration~\cite{Abeysekara:2010ze,Allen:2009aa,Cortese:2008zza}.  The EMCal had 11520 towers with transverse size 6 cm $\times$ 6 cm, or approximately twice the effective Moli\`ere radius.  The relative energy resolution is $0.11/\sqrt{E}+0.017+0.051/E$, where the energy $E$ is measured in GeV~\cite{Allen:2009aa}.  Clusters are formed by combining signals from adjacent towers and each cluster is required to have only one local energy maximum.  This analysis uses events triggered by a high energy deposit in a 4$\times$4 region of towers in the EMCal.  This trigger configuration has less sensitivity to the underlying event than a trigger configuration with a larger area, often used for jet analyses, because the contribution of the underlying event to the energy is proportional to the trigger area.  The raw trigger threshold was multiplicity dependent and corresponded to approximately 4.5--6~GeV in the centrality bin used in this analysis.

Clusters with energy above 3 GeV, which exclude minimally ionizing particles~\cite{Cortese:2008zza}, are used in this analysis.  Partially formed hadronic showers may still pass this threshold.  To avoid overcounting of charged particle \pt, the cluster energies are corrected as in~\cite{Abelev:2013fn}.  Tracks are propagated to the average cluster depth in the EMCal and matched to the nearest cluster.  If the nearest cluster is within $|\Delta \eta|< 0.015$ in pseudorapidity and $|\Delta \phi|< 0.025$ in azimuth, the cluster most likely arose from a charged hadron.  If the cluster energy is at or below the track's momentum, the cluster is not used in the analysis, while if the cluster energy is above the track's momentum, the track momentum is subtracted from the cluster energy~\cite{Abelev:2013fn}.

\section{Method}\label{Sec:Methods}
\begin{figure}[!htbp]
\begin{center}
\rotatebox{0}{\resizebox{5cm}{!}{
        \includegraphics{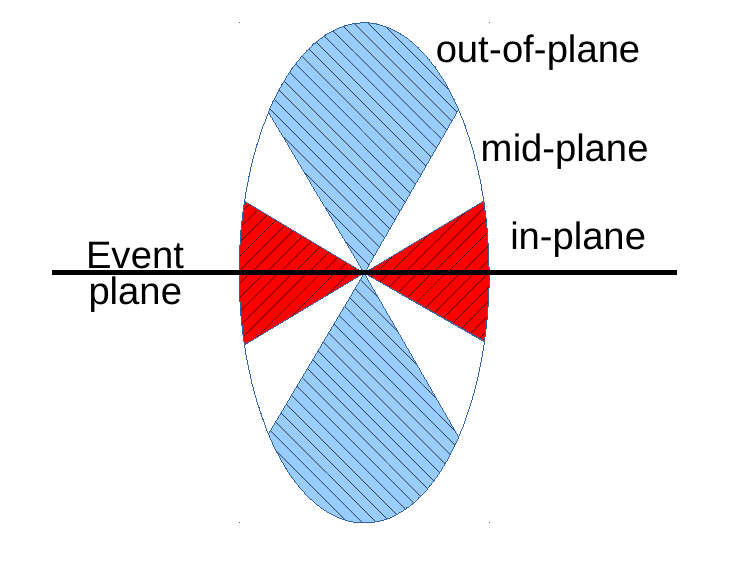}
}}
\caption{Jet-hadron correlations are measured for jets in three regions relative to the $n = 2$ event plane, which is transverse to the direction of the beams.  These regions include in-plane ($| \psi-\phi_{\mathrm {jet}}|<\pi/6$) shown in red, mid-plane ($\pi/6<|\psi-\phi_{\mathrm {jet}}|<\pi/3$) shown in white, and out-of-plane ($|\psi-\phi_{\mathrm {jet}}|>\pi/3$) shown in blue.
}\label{plot:rxnschematic}
\end{center}
\end{figure}

The data used in this analysis were collected during the 2011 run~\cite{Abelev:2014ffa} from 0.5M 30--50\% central \PbPb collisions at \sNN = 2.76 TeV.  It was additionally required to be triggered by a high energy deposit in a 4$\times$4 region of towers in the EMCal~\cite{Cortese:2008zza}.  Procedures for selection and reconstruction of tracks from charged particles, identification of calorimeter clusters, and jet reconstruction are as in~\cite{Adam:2015ewa} and are summarized in \sref{Sec:JetReconstruction}.  Estimates of the distributions of corrected jet energies are also reported here.  The experimentally reconstructed second order symmetry plane is called the second order event plane, referred to as the ``event plane" later in the text for simplicity.  Centrality determination and event plane reconstruction is discussed in \sref{Sec:CentralityEventPlaneDet}.  Jets from triggered events are correlated with all charged tracks in azimuth ($\Delta\phi = \phi_{\mathrm{jet}}-\phi_{\mathrm{assoc}}$) and pseudorapidity ($\Delta\eta = \eta_{\mathrm {jet}}-\eta_{\mathrm {assoc}}$), as discussed in \sref{Sec:Correlations}.  The distributions of these associated tracks relative to the trigger jet are measured in three bins in the angle between the trigger jet and the event plane, in-plane ($| \psi-\phi_{\mathrm {jet}}|<\pi/6$), mid-plane ($\pi/6<|\psi-\phi_{\mathrm {jet}}|<\pi/3$), and out-of-plane ($|\psi-\phi_{\mathrm {jet}}|>\pi/3$) bins, as shown in \fref{plot:rxnschematic}.  The analysis is restricted to 30--50\% central \PbPb collisions because this is where the event plane resolution is highest and therefore the analysis will be most sensitive to any path length dependencies.  The subtraction of the combinatorial background using the RPF method~\cite{Sharma:2015qra} is discussed in \sref{Sec:BgSub}.  The determination of the yields and the widths is discussed in \sref{Sec:YieldAndWidth}.  The possible impact of the finite event plane resolution on the signal is discussed in \sref{Sec:RxnPlaneResolutionEffects}.

\subsection{Jet reconstruction and energy distribution}\label{Sec:JetReconstruction}
Tracks and corrected EMCal clusters are clustered into jets using the \akT algorithm with a resolution parameter $R = 0.2$ in the FastJet package~\cite{Cacciari:2011ma}.  Jet transverse momenta are calculated as the scalar sum of their constituent transverse momenta using a boost-invariant \pt recombination scheme.  Tracks are assumed to be pions and clusters to arise from massless particles.  In order to suppress contributions from combinatorial jets and the contribution of uncorrelated background to the jet energy, jets are reconstructed with constituents above \ptabove{3} and are required to have an area of at least $0.08$ calculated using ghost particles as described in~\cite{Cacciari:2011ma}.  Jets are required to be within $|\eta_{\mathrm{jet}}| < 0.5$ and $ 1.6 < \phi_{\mathrm{jet}} < 2.9$ so that the entire jet is contained within the EMCal acceptance.  

Small jets are used to reduce the impact of the background, as background contributions scale with $R^2$.  Additionally, to further suppress contributions of the background to the energy and to match trigger  conditions~\cite{Abeysekara:2010ze}, the jets are required to contain a cluster with transverse energy larger than 6~GeV.  We note that this requirement leads to a selection of biased jets, explicitly biasing the \ns.  The \as, in contrast, is not explicitly biased, although it is unlikely to be a random sample of the jet population.  With these constituent cuts, the background contribution to the energy is negligible using estimates of the background per unit area as in~\cite{Abelev:2012ej}.  The background contribution to the energy is therefore not subtracted from the jets, although any residual contribution would be included in the energy distribution estimation.  The jet energy is not corrected for detector effects, but the distribution of particle level jet energies in the sample is estimated using {\sc pythia6}~\cite{Sjostrand:2006za} Tune A~\cite{Field:2005sa} simulations embedded at detector level into data measured in \PbPb collisions and matched back to generator level.

Detector effects such as the single track reconstruction efficiency, momentum resolution in the tracking 
detectors and energy resolution in the calorimeter combined with contributions from particles which are 
not directly observed, such as neutrons and $K^0_{\mathrm L}$, and contributions from the background lead to a finite 
energy resolution.  This means that when jet-hadron correlations are measured at a particular jet \pt, the 
distribution of true jet momenta is broad.  A full correction for this effect would require two dimensional 
unfolding with jet-hadron correlations measured for several jet momenta.  The current statistics do not allow 
for such measurements.  Instead, we estimate the distribution of true jet energies and focus on comparisons between jets at different angles relative to the event plane to enable the highest precision search for path length dependence allowed by the currently available data.

The kinematic selection of tracks and clusters used in jet finding is chosen to suppress contributions from the combinatorial background and reduce smearing of the jet energy due to the large combinatorial background.  
With these kinematic selections, the background density is negligible and showed no event plane dependence.  Given that no event plane dependence is observed in the signal, this also means that the resolution of the jet axis does not vary with the angle of the jet relative to the event plane.

\begin{figure}[!htbp]
\begin{center}
\rotatebox{0}{\resizebox{10cm}{!}{
        \includegraphics{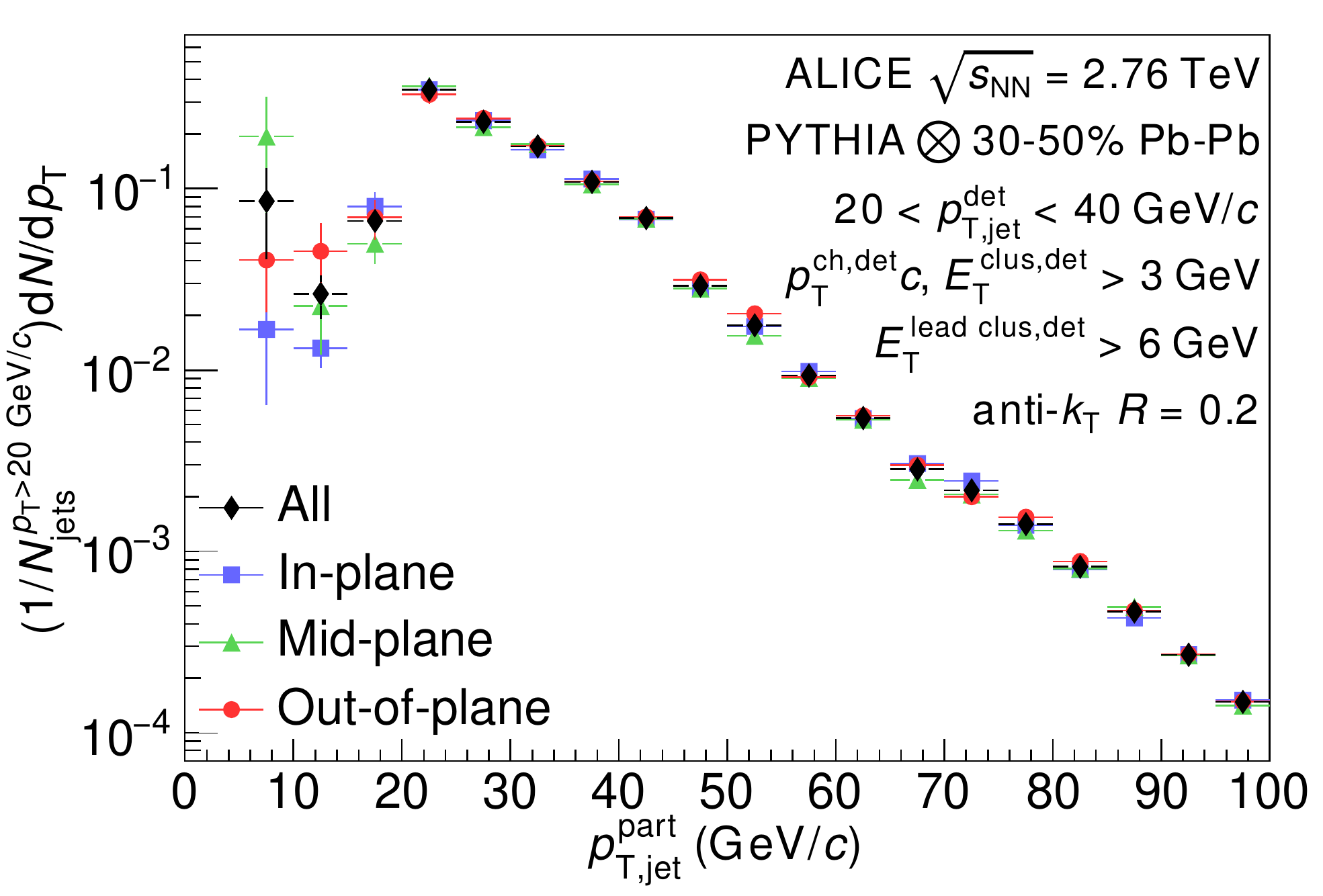}
}}
\caption{The generator level jet probability distribution corresponding to jets measured with \pt between 20-40 \GeV 
for PYTHIA events embedded in 30--50\% central \PbPb collisions.  Generator level jets are required to have $\pt>5$ \GeV.  The distribution is measured
for each angle relative to the event plane, as well as the sum of all angles.  
}\label{plot:exampleParticleJetDistribution}
\end{center}
\end{figure}

{\sc pythia6} TuneA~\cite{Sjostrand:2006za,Field:2005sa} simulations of \pp collisions with jets
are embedded at detector level into 30-50\% \PbPb data.  The embedded events are analyzed with
the same parameters and cuts as the data analysis, including the jet constituent
and cluster biases, while the generator level jets are measured for $\pt>5$ \GeV.  The generator level jets are
first matched geometrically to {\sc PYTHIA} detector level jets, and then the detector level jets are
geometrically matched to jets found within $R=0.2$ in the embedded event, with the additional requirement
the associated generator level jet distribution is measured.  Each such distribution is normalized within the region $20 \leq \pt < 100$ \GeV where fluctuations are minimized
and, assuming that the jet energy distribution in the data is the same as that provided by {\sc PYTHIA6}, describes the
generator level jet distribution that corresponds to the measured detector level jet distribution, shown 
in \fref{plot:exampleParticleJetDistribution}.  The uncertainty on the jet energy scale is 2.6\% and the jet energy resolution, which is also encoded in the response matrix, is around 20\% for the jets selected in this analysis~\cite{Abelev:2013fn}.
There are slight differences in the jets reconstructed with $\pt \leq 20$ \GeV for jets at different angles relative to the event plane due 
to a low momentum embedded jet overlapping 
with a another jet in the \PbPb data.  Since there are more jets in-plane than out-of-plane in the data, this leads to an apparent difference in the reconstructed jet spectra.  Otherwise there are no significant differences between 
jets at different angles relative to the event plane.

\subsection{Centrality determination and event plane reconstruction}\label{Sec:CentralityEventPlaneDet}
Centrality is determined from the sum of the energy deposition~\cite{Abbas:2013taa} in the V0 scintillator tiles, as described in~\cite{Abelev:2013qoq}.  The centrality of the collision is reported as percent ranges of the total hadronic cross section, with lower percentiles referring to the most central (largest multiplicity) events~\cite{Abelev:2013qoq}.  
The second order event plane $\Psi_{\mathrm {EP},2}$ is reconstructed using the V0 following the procedure in~\cite{Adam:2015mda} by combining signals from the V0A and V0C arrays~\cite{Cortese:2004aa}.  The separation in pseudorapidity between the measurement of the signal and the measurement of the event plane suppresses the contribution from the jet signal to the event plane determination~\cite{Adam:2015mda}.

The reaction plane is defined by the beam axis and the vector between the centers of the two incoming nuclei.  Additional asymmetries in the distribution of nucleons within the overlap region, generally quantified by a harmonic decomposition, generate symmetry planes at all orders ($n > 0$)~\cite{Sorensen:2010zq,Alver:2010gr}.  If the nucleons were in their average positions and there were no fluctuations in interactions between nucleons, the reaction plane would correspond to the second order symmetry plane, $\Psi_{EP,2}$.  The experimentally reconstructed symmetry planes are called event planes.  For simplicity, we refer here to the experimentally reconstructed second order event plane as the event plane.  As explained below, we do not correct the signal for the difference between the event and symmetry planes because no event plane dependence is observed in this measurement.  Corrections for the event plane resolution will increase differences between results at different angles relative to the event plane but will not induce any event
plane dependence if there is none in the uncorrected results.
We discuss the impact of the event plane resolution in \sref{Sec:RxnPlaneResolutionEffects}.  The impact of the dijets on the event plane reconstruction was studied in~\cite{Adam:2015mda} and found to be negligible.

The event plane is also extracted using TPC tracks in order to determine the event plane resolution.  The $n$-th order event plane can be calculated from the charged particle azimuthal distribution by~\cite{Poskanzer:1998yz}:
\begin{equation}
 \Psi_{n,EP} = \left( \arctan \frac{\sum_{i} w_i \sin(n\phi_i)}{\sum_{i} w_i \cos(n\phi_i)} \right) /n ,
\end{equation}
\noindent where the sum is over all particles in the event, $\phi_i$ is the azimuthal angle of the $i$-th particle, and $w_i$ is the weight of the $i$-th particle.  For measurement of the event plane with the V0, the sum is over all of its sectors and the weights are equal to the amplitude of the respective sector in the V0, which is proportional to the local multiplicity.  A calibration and recentering procedure following ~\cite{Adam:2015mda} is applied to remove any bias introduced by non-uniform acceptance of the V0 system.  For measurement in the TPC, tracks are given equal weights ($w_i$ = 1) and the acceptance is nearly uniform.  Track selection is the same as that described in \sref{Sec:Methods} except the transverse momentum range is restricted to 0.15~$< \pt < $~5.0~\GeV.  More details of the procedure can be found in~\cite{Adam:2015mda}.

Due to the finite multiplicity of each event, there will be a difference between the symmetry plane and the reconstructed event plane. This difference is quantified by the event plane resolution
\begin{equation}
  \Re_{n} = \langle \cos(n[\Psi_{\mathrm {n,EP}} - \Psi_n]) \rangle.
  \label{resolution}
\end{equation}
where $\Psi_{n}$ is the true angle and $\Psi_{n,EP}$ is the measured angle of the $n$th order event plane.  To evaluate the event plane resolution directly from data, this analysis uses three sub-events, comparing the event planes measured in the full V0, measured in the TPC using tracks at positive pseudorapidities, and measured in the TPC using tracks at negative pseudorapidities.  We can express the $n$th order resolution of the full V0, $\Re^{V0}_{n}$~\cite{Barrette:1996rs,Poskanzer:1998yz}, of the second order event plane by
\begin{equation}
  \Re^{\mathrm{V0}}_{n} = \langle \cos(n[\Psi^{\mathrm{V0}}_{2,\mathrm{EP}} - \Psi_2]) \rangle = \sqrt{\frac{\langle \cos(n[\Psi^{V0}_{2,\mathrm{EP}} - 
  \Psi^{\mathrm{TPC},\eta>0}_{2,\mathrm{EP}}]) \rangle \langle \cos(n[\Psi^{\mathrm{V0}}_{2,\mathrm{EP}} - \Psi^{\mathrm{TPC},\eta<0}_{2,\mathrm{EP}}])\rangle }{\langle 
  \cos(n[\Psi^{\mathrm{TPC},\eta>0}_{2,\mathrm{EP}} - \Psi^{\mathrm{TPC},\eta<0}_{2,\mathrm{EP}}])\rangle }}
  \label{RESV0}
\end{equation}
\noindent where $\Psi^{V0}_{2,\mathrm{EP}}$, $\Psi^{\mathrm{TPC},\eta>0}_{2,\mathrm{EP}}$, and $\Psi^{\mathrm{TPC},\eta<0}_{2,\mathrm{EP}}$ are the second order event planes calculated using the three different sub-events, and $\Psi_2$ is the true angle of the second order symmetry plane.  Fits to the Fourier decomposition of the correlated background are performed up to $n = 4$ and are measured relative to the event plane.  Thus, event plane resolution corrections $\Re_{2}(\psi_{2})$ and $\Re_{4}(\psi_{2})$ are needed to correct these terms for the finite precision of the second order event plane measured in the V0 system, as discussed in \sref{Sec:BgSub}.  The event plane resolution for the 30-40\% and 40-50\% centrality ranges are combined by weighting the two samples by the number of corresponding events of each.  The event plane resolutions $\Re_{2}(\psi_{2})$ and $\Re_{4}(\psi_{2})$ for the analyzed 30-50\% event sample are 0.73 and 0.44, respectively, with negligible uncertainties.

\subsection{Jet-hadron correlations}\label{Sec:Correlations}
The distribution of charged particles relative to \ptjetrange{20}{40} reconstructed jets is measured in azimuth (\dphi) and pseudorapidity (\deta) as 
\begin{equation}
 \frac{1}{N_{\mathrm{trig}}} \frac{d^2N_{\mathrm{assoc}}}{d\Delta\phi d\Delta\eta} = \frac{1}{N_{\mathrm{trig}}} \frac{1}{ a(\Delta\phi,\Delta\eta) \epsilon(p_{\rm T}^{\mathrm{assoc}},\eta^{\mathrm{assoc}})} \frac{d^2 N_{\mathrm{meas}} }{d\Delta\phi d\Delta\eta} 
  \label{Eqn:correlationfncwithbkgd}
\end{equation}
\noindent where $N_{\mathrm{trig}}$ is the number of trigger jets, $\epsilon(p_{\rm T}^{\mathrm{assoc}},\eta^{\mathrm{assoc}})$ is the product of the single track reconstruction efficiency and acceptance, and $a(\Delta\phi,\Delta\eta)$ dominantly corrects for the pair acceptance.  The distributions are determined in bins of centrality, associated hadron transverse momentum ($\it{p}_{\rm T}^{\mathrm{assoc}}$), and bins of the trigger jet angle relative to the event plane.

The correction $a(\Delta\phi,\Delta\eta)$ is calculated as a function of centrality and associated particle momentum by mixed events using a trigger jet from an EMCal-triggered event and associated hadrons from minimum bias events or semi-central triggered events.  The mixed event procedure will also remove the trivial correlation due to an $\eta$ dependence in the single particle and track distributions.  However, since there is little $\eta$ dependence in either tracks or jets within the acceptance used in this analysis, the dominant effect is the pair acceptance.  Mixed events are constructed separately for 30--40\% and 40--50\% centrality classes.  The mixed events are required to be within the same 10\% centrality class and have vertex positions within 2 cm along the direction of the beam, $z_{\mathrm{vtx}}$.  There is no difference in the correction within uncertainties for different orientations of the jet relative to the event plane, and therefore the same correction $a(\Delta\phi,\Delta\eta)$ is applied for all angles relative to the event plane.  All associated momentum bins for \ptabove{2.0} are combined to increase statistics because $a(\Delta\phi,\Delta\eta)$ has little momentum dependence at high momenta.  The correction $a(\Delta\phi,\Delta\eta)$ is normalized to one at its maximum with the systematic uncertainty in the normalization determined by using different regions in \dphi and \deta, with a systematic uncertainty below 0.5\% for all $\it{p}_{\rm T}^{\mathrm{assoc}}$ bins used in this analysis.  There is an additional shape uncertainty due to slight changes in the correlation function at large \deta in the acceptance with $z_{\mathrm {vtx}}$ position.  Since the background level is determined from the level of the correlation function at large \deta, this leads to a scale uncertainty in the background subtraction.  This uncertainty is determined by varying the binning of the mixed events in $z_{\mathrm{vtx}}$ and is correlated for different angles relative to the event plane and for different bins in \ptassoc.  This scale uncertainty is used later for determining a systematic uncertainty on the background subtraction and is dependent on \ptassoc.  

\subsection{Background subtraction}\label{Sec:BgSub}
The signal in \eref{Eqn:correlationfncwithbkgd} has a large combinatorial background from particles created by processes other than the hard process which created the jet.  The jet signal may be correlated with the second order event plane because of jet quenching, and soft hadrons are correlated with the second order event plane due to hydrodynamical flow.  The Fourier expansion of this background can be expressed by:

\begin{equation}
 \frac{dN}{\pi d\Delta\phi} = B \bigg( 1 + \sum_{n = 1}^{\infty} 2 \tilde{v}_{n}^{\mathrm{trig}} \tilde{v}_{n}^{\mathrm{assoc}} \cos(n\Delta\phi) \bigg), \label{Eqtn:JBBBCorrelations}
\end{equation}

where $\tilde{v}_{n}^{\mathrm{trig}}$ and $\tilde{v}_{n}^{\mathrm{assoc}}$ refer to the effective Fourier coefficients for the azimuthal anisotropy of the trigger jet and associated hadron, respectively, to the background.  For inclusive measurements, if the background is dominantly due to flow, the $\tilde{v}_{n}$ of this background will be equal to the $v_{n}$ due to flow.  The exact values may be slightly different due to differences in the event samples, varying sensitivity in the method to fluctuations in the $v_n$ and non-flow, the difference between the average over all pairs $\langle \tilde{v}_{n}^{\mathrm{trig}} \tilde{v}_{n}^{\mathrm{assoc}} \rangle$ and the product of the averages over all events $\langle \tilde{v}_{n}^{\mathrm{trig}} \rangle \langle \tilde{v}_{n}^{\mathrm{assoc}} \rangle$, differences in the $v_n$ for particles in jets and from the bulk, and decorrelations between symmetry planes for hard and soft processes.  

The contribution from these soft processes is subtracted using the RPF method~\cite{Sharma:2015qra}.  This method avoids contamination by the near- and away-side jets by focusing on the \ns only at large \deta and instead using the dependence of the flow-modulated background on the angle of the trigger jet relative to the event plane to constrain the background shape and level.  For in-plane jets, background particles are more likely to be near the trigger jet than $\pi$ away in azimuth, leading to a higher $\cos(2\Delta\phi)$ term, and the background level is higher because there are more jets in-plane.  For out-of-plane jets, background particles are less likely to be near the trigger jet, leading to a negative $\cos(2\Delta\phi)$ term, and the background is lower because there are fewer jets.  Because the second and fourth order event planes are correlated, a similar argument holds for the fourth order terms.  These effects help constrain the even $n$ terms and help distinguish them from the odd $n$ terms and constrain the background level while avoiding contamination from the near- and away-side jets.

The event plane dependence can be used to determine the background shape and level.  When the angle of the jet is fixed relative to the event plane, the effective size and shape of the background is given by

\begin{center}
\begin{align}
 \tilde{B} & = \frac{N_t N_a  c}{\pi^2} \Big(1
 +  2 \sum_{k = 1}^{\infty}  \frac{v_{2k}^{\mathrm{trig}}  }{2kc} \sin(2kc) \Re_{2k} C_{2k,0} \cos(2k\phi_s) \Big), \label{Eq:BFE} \\
 \tilde{v}_{n}^{\mathrm{trig}} &= \frac{ v_{n} + \frac{\delta_{n,\mathrm{mult}\ 2} }{nc} \sin(nc) \Re_{n} C_{n,0}   \cos(n\phi_s)+ \sum_{k = 1}^{\infty} (v_{2k+n}^{\mathrm{trig}}C_{|2k+n|,n}+v_{|2k-n|}^{\mathrm{trig}}C_{|2k-n|,n}) \frac{\sin(2kc)\cos(2k\phi_s) \Re_{2k}}{2kc}}{
 1 +  2 \sum_{k = 1}^{\infty}  \frac{v_{2k}^{\mathrm{trig}}  }{2kc} \sin(nc) \Re_{2k} C_{2k,0} \cos(2k\phi_s)
 } \nonumber \\ 
 C_{n,m} & = \langle \cos(n\psi_{n}+m\psi_{m}-(n+m)\psi_{2}) \rangle \nonumber
\end{align}
\end{center}

where $\phi_s$ is the center of the azimuthal range of the trigger particle relative to the event plane, $c$ is the width of that range, $N_t$ is the number of triggers, $N_a$ is the number of associated particles, $v_{n}^{\mathrm{assoc}}$ are the \vn of the associated particles, $v_{n}^{\mathrm{trig}}$ are the \vn of the triggers, and $\Re_{n}$ is the event plane resolution given in \eref{RESV0}~\cite{Bielcikova:2003ku,Nattrass:2018jzg}.  Terms \vn with $n<1$ are zero.  The $C_{n,m}$ terms are a measure of how correlated event planes of different orders are with the second order event plane and are approximately zero when either $n$ or $m$ is odd.  This is consistent with the weak correlation between the $n = 2$ participant plane and odd order participant planes~\cite{Aad:2014fla} because the odd $\tilde{v}_{n}$ arise mainly due to fluctuations in the initial state. In this case, the even $\tilde{v}_{n}$ will change when the angle of the jet relative to the event plane is changed while the odd $\tilde{v}_{n}$ remain constant.  The equation is expanded to include terms up to \vnum{4}.  The term $C_{2,0}$ = 1 and the terms $C_{4,0}$ and $C_{4,2}$ are approximated to be one.  The latter assumption will lead to an inconsistency between the \vnum{4} from independent measurements and from the fit, but the fit will still provide a valid description of the background.

The shape of the background depends on the $\Re_{n}$, which are fixed at the measured values.  The fourth order event plane resolution is calculated relative to the second order event plane, consistent with the shape described in \eref{Eq:BFE}.  The uncertainties on the event plane resolution are negligible relative to the statistical and background fit uncertainties of the final results.  

The jet signal in \eref{Eqn:correlationfncwithbkgd} can be decomposed into a \ns and an \as.  The \ns is a peak which is narrow in both \dphi and \deta, meaning that it is negligible at large \deta, while the \as is narrow in \dphi but broad in \deta.  The correlation function at large \deta ($0.8<|\Delta\eta|<1.2$) and small \dphi ($|\Delta\phi|<\pi/2$) is fit simultaneously for $\tilde{v}_n$ up to $n = 4$ for trigger jets in-plane ($| \psi-\phi_{\mathrm jet}|<\pi/6$), mid-plane ($\pi/6<|\psi-\phi_{\mathrm jet}|<\pi/3$), and out-of-plane ($|\psi-\phi_{\mathrm jet}|>\pi/3$), shown in \fref{plot:rxnschematic}, to determine the background shape and level.  Because the even $\tilde{v}_n$ depend on the angle of the jet relative to the event plane, as shown in \eref{Eq:BFE}, the $\tilde{v}_2$ and $\tilde{v}_4$ of both the trigger jet and the associated particle are determined in the fit while only the product $\tilde{v}_{3}^{\mathrm {jet}} \tilde{v}_3^{\mathrm {assoc}}$ is extracted from the fit.  A rapidity-even $\tilde{v}_1$ term can arise due to both momentum conservation and fluctuations in the initial state.  This rapidity-even $\tilde{v}_1$ has been measured for single hadrons and is comparable in magnitude to $\tilde{v}_2$ and $\tilde{v}_3$~\cite{Luzum:2010fb,ATLAS:2012at}.  This term does not change when the angle of the trigger jet is varied relative to the event plane so the product $\tilde{v}_{1}^{\mathrm {jet}} \tilde{v}_1^{\mathrm {assoc}}$ contributes to the background.  When the fit function is varied to include this $n = 1$ term, it is zero within uncertainties and did not lead to significant differences in the correlation function.  Since $\tilde{v}_1^{\mathrm {assoc}}$ is known to be non-zero, this likely means that $\tilde{v}_{1}^{\mathrm {jet}}$ is near zero.  For associated particles above \ptabove{2}, the background is low and the statistics for the region which is background-dominated on the \ns are therefore also low, so the fit is restricted up to $n = 3$.  The fits used in this analysis therefore have six parameters below 2~\GeV, $B$, $\tilde{v}_{2}^{\rm jet}$, $\tilde{v}_{2}^{\mathrm {assoc}}$, $\tilde{v}_{3}^{\mathrm {jet}} \tilde{v}_3^{\mathrm {assoc}}$, $\tilde{v}_{4}^{\mathrm {jet}}$, and $\tilde{v}_{4}^{\rm assoc}$, and four above.  The event plane resolution is fixed and variations within the uncertainties lead to negligible differences in the correlation functions.  

\begin{figure}[!htbp]
\begin{center}
\rotatebox{0}{\resizebox{\textwidth}{!}{
        \includegraphics{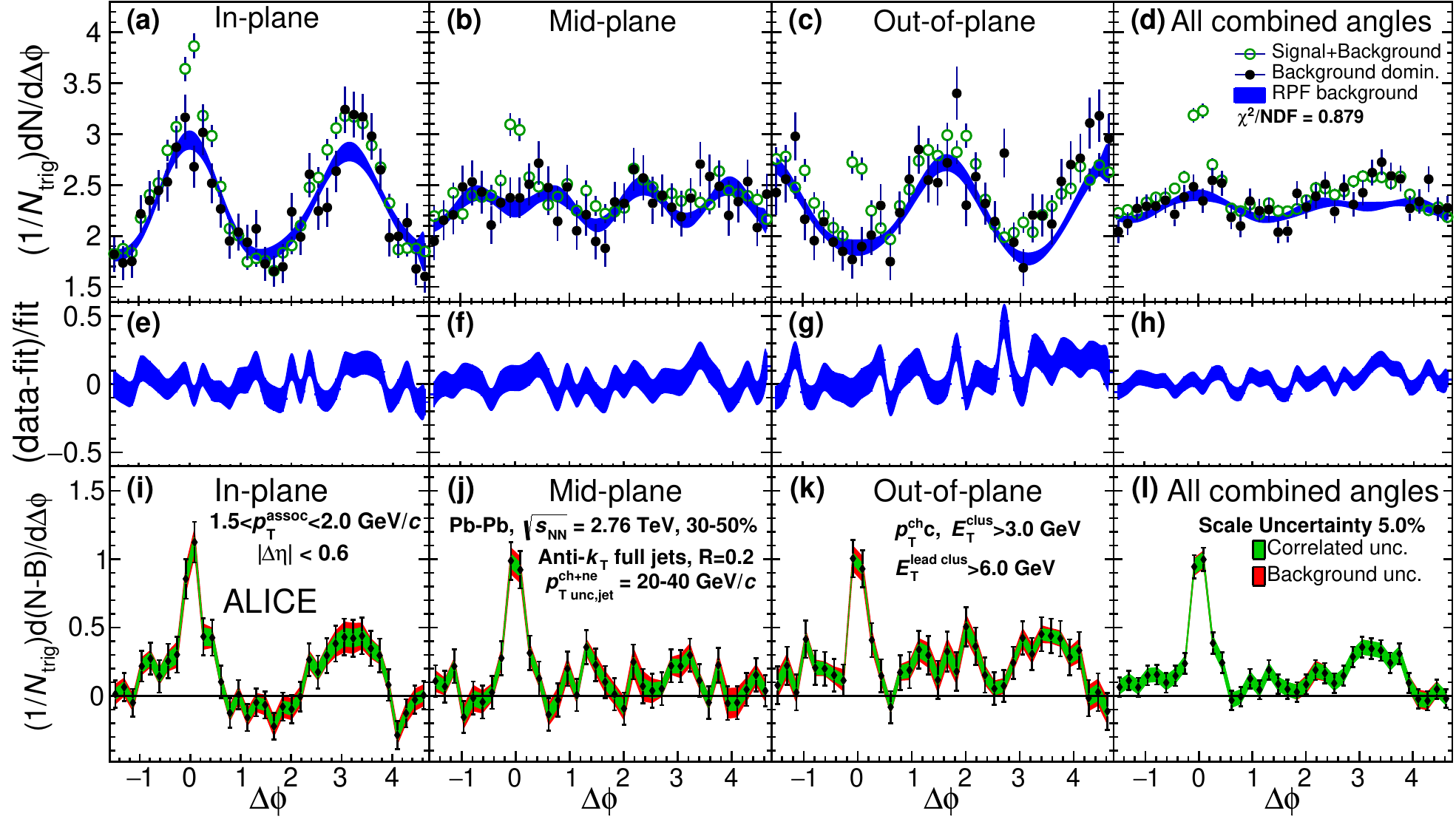}
}}
\caption{The signal plus background region, $|\Delta\eta|<0.6$ (green points), the region which is background-dominated on the \ns, 
  $0.8<|\Delta\eta|<1.2$ (black points), and the RPF fit to $|\Delta\phi|<\pi/2$ (blue band) to the region which is background-dominated on the \ns
  for \pttrigrange{20}{40} jets correlated with \ptassocrange{1.5}{2.0} hadrons from 30-50\% centrality collisions on the 
  top panel.  The middle panel shows the quality of the RPF fit to the region which is background-dominated on the \ns, (data$-$fit)/fit.  
  On the bottom panel are the RPF corrected correlation functions, with the uncertainty from the background fit (red band), 
  and the correlated uncertainty (green band).
}\label{plot:dPhi1.5-2.0}
\end{center}
\end{figure}

\Fref{plot:dPhi1.5-2.0} shows a sample correlation function in the region which is background-dominated on the \ns compared to the fit for in-plane (a), mid-plane (b), and out-of-plane (c) jets and all jets combined (d) for associated particles with momenta \ptassocrange{1.5}{2.0}.  Correlation functions for other \ptassoc ranges used in this analysis are given in~\cite{ALICE-PUBLIC-2019-004}.  The ratio of the difference between the data and the fit to the fit is shown in \fref{plot:dPhi1.5-2.0}(e-h), showing that the fit describes the data well.  The background subtracted correlation functions in the region $|\Delta\eta|<0.6$ are shown in \fref{plot:dPhi1.5-2.0}(i-l).  The uncertainties from the background subtraction are propagated using the covariance matrix from the fit.  These uncertainties are non-trivially correlated point-to-point and between different bins relative to the event plane and are shown as a blue band.  The scale uncertainties on the background are shown as a green band and are correlated point-to-point and between different bins relative to the event plane.  The uncertainties due to the single track reconstruction efficiency, contamination, and the normalization of the acceptance correction are uncorrelated with each other but correlated for all points.  These uncertainties are combined and listed as the scale uncertainty.  The $\tilde{v}_{2}^{\mathrm{jet}}$ and $\tilde{v}_{2}^{\mathrm{assoc}}$ extracted from the fits are in agreement with other ALICE results~\cite{Adam:2015mda,Abelev:2012di}.  

Finite event plane resolution reduces the event plane dependence of the signal because the measurement in one bin relative to the event plane will have contributions from other bins as well.  Techniques for correcting for this effect increase the event plane dependence observed in the uncorrected data~\cite{Adare:2018wjb}, but if there is no event plane dependence in the uncorrected data, these corrections will not reveal an event plane dependence.  Since no event plane dependence is observed within uncertainties and corrections would increase the complexity of the measurement and the systematic uncertainties, no correction is applied for the finite event plane resolution.  The impact of the finite event plane resolution is discussed in \sref{Sec:RxnPlaneResolutionEffects}.

\subsection{Associated track yields and peak widths}\label{Sec:YieldAndWidth}

The yield of tracks associated with jets is calculated by integrating the associated yield:

\begin{equation}
 Y = \frac{1}{N_{\mathrm{trig}}} \int_{\eta_1}^{\eta_2} \int_{\phi_1}^{\phi_2} \frac{d(N_{meas} - N_{bkgd})}{d\Delta\phi}  d\Delta\phi d\Delta\eta. \label{Eqn:Yield2}
\end{equation}
\noindent The integration limits of $\phi_1 = -\pi/3$ and $\phi_2 = \pi/3$ for the \ns, $\phi_1 = 2\pi/3$ and $\phi_2 = 4\pi/3$ for 
the \as, and $\eta_1 = -0.6$ and $\eta_2 = 0.6$ for both are part of the definition of the measurement.  The systematic uncertainties due to the extraction of the background from the RPF are propagated using the covariance matrix from the fit.  This uncertainty is non-trivially correlated between yields for different angles of the jet relative to the event plane and for the near- and \as and uncorrelated for points at different \ptassoc.  The shape uncertainty in the acceptance correction at large \deta leads to an additional scale uncertainty when propagating the background determined in the region $0.8<|\Delta\eta|<1.2$ to the region ($|\Delta\eta|<0.6$).  This uncertainty is 100\% correlated for all data points.  The single track reconstruction efficiency uncertainty, the uncertainty due to normalization of the acceptance correction, and the uncertainty due to contamination from secondary particles are 100\% correlated for all points and affect the scale of the correlation functions and the yields.

The ratios and differences of yields are calculated in order to investigate possible event plane dependent modifications.  The systematic uncertainties from the background largely cancel out in the ratio and the difference.  The track reconstruction efficiency, mixed event normalization, and secondary contamination systematic uncertainties cancel out in the ratio.

The widths are quantified by fitting the correlation functions to a Gaussian, $A \rm e^{(\Delta\phi - \Delta\phi_0)^2/2\sigma^2}$ where $\Delta\phi_0~=~0$ on the \ns and $\Delta\phi_0~=~\pi$ on the \as, in the range $|\Delta\phi|~<~\pi/3$ on the \ns and $|\Delta\phi - \pi|~<~\pi/3$ on the \as.  The near- and \as are fit separately.  The Gaussian fit is repeated with different values of the background parameters and the covariance matrix is used to propagate the uncertainties.

The systematic uncertainties are summarized in \tref{Tab:SystematicSummary} and \tref{Tab:SystematicSummaryEPdep}.  \Tref{Tab:SystematicSummary} lists the sources of systematic uncertainties which are independent of the angle relative to the event plane and the momentum, including the single track reconstruction efficiency (\sref{Sec:Detector}), contamination from secondaries (\sref{Sec:Detector}), uncertainties in the mixed events due to their normalization and shape in \dphi (\sref{Sec:Correlations}), and uncertainties in the event plane resolution (\sref{Sec:CentralityEventPlaneDet}).  The uncertainties in the single track reconstruction efficiency, normalization of the acceptance correction determined from mixed events, and secondary contamination lead to a 5\% uncertainty in the scale of the correlation functions and yields.  This uncertainty is uncorrelated for different associated particle momenta.

\Tref{Tab:SystematicSummaryEPdep} lists uncertainties which are dependent on the angle of the jet relative to the event plane and the associated particle's momentum on the yields due to the scale uncertainty in the mixed events (\sref{Sec:Correlations}) and in the background fit (\sref{Sec:BgSub}) for a few representative associated particle momenta.  The uncertainty of the acceptance correction determined from mixed events in \deta and the uncertainty due to the background subtraction are different for different \ptassoc bins and therefore shown separately for each data point.  The uncertainty due to the shape uncertainty of the acceptance correction determined from mixed events in \deta is correlated for different angles relative to the event plane and uncorrelated between different \ptassoc bins.  The uncertainty due to the background subtraction is non-trivially correlated for different angles of the jet relative to the event plane but uncorrelated between different \ptassoc bins.  

\begin{table}[!htbp]
\caption{Summary of systematic uncertainties which are independent of the angle relative to the event plane and the momentum for \pttrigrange{20}{40} in 30-50$\%$ central \PbPb collisions.}\label{Tab:SystematicSummary}
\centering 
\begin{tabular}{l | c} 
\hline 
Source & Uncertainty $\%$ \Tstrut \Bstrut \\ 
\hline 
Single particle reconstruction efficiency & 4 \\
\hline
Contamination & 1 \\
\hline
Mixed event (shape $\Delta\phi$) & negligible \\
\hline
Mixed event normalization & $<$ 0.5 \\ 
\hline
event plane resolution & negligible \\ 
\hline
\end{tabular}
\end{table}

\begin{table}[!htbp]
\caption{Summary of systematic uncertainties on the yields and widths calculated from the correlation functions due to the shape uncertainty coming from the shape of the acceptance correction in \deta and the correlated background fit uncertainty, both varying with event plane orientation bins.  They are displayed for \pttrigrange{20}{40} in 30-50$\%$ central \PbPb collisions for \ptassocrange{1.0}{1.5} and \ptassocrange{3.0}{4.0} bins.  The values are expressed as a percent of the nominal value.}
\label{Tab:SystematicSummaryEPdep}
\centering
\begin{tabular}{c | c | c | c | c | c | c}\hline
  \multirow{3}{*}{Source} & \multirow{3}{*}{Result} & \multirow{3}{*}{Orientation} & \multicolumn{4}{c}{Uncertainty $\%$} \\ \cline{4-7}
  &   &      & \multicolumn{2}{c|}{Near-side: $\it{p}_{\rm_T}^{\mathrm{assoc}}$ (\GeV)} & \multicolumn{2}{c}{Away-side: $\it{p}_{\rm_T}^{\mathrm{assoc}}$ (\GeV)} \Bstrut \\ \cline{4-7}     
  &   &      & 1.0-1.5 & 3.0-4.0 & 1.0-1.5 & 3.0-4.0 \\ \hline                        
  \multirow{6}{*}{} & \multirow{3}{*}{Yield} & in-plane & 20 & 2.8 & 33 & 7.9  \\ \cline{3-7}
  & & mid-plane & 13 & 2.7 & 25 & 9.2 \\ \cline{3-7}
  Acceptance & & out-of-plane & 10 & 2.5 & 22 & 6.3 \\ \cline{2-7}
  shape & \multirow{3}{*}{Width} & in-plane & 14 & 1.5 & - & 5.0 \\ \cline{3-7}
  & & mid-plane & 9.8  & 1.4 & - & 7.1 \\ \cline{3-7}
  & & out-of-plane & 5.9 & 0.9 & - & 4.6 \\ \cline{3-7}
  \hline
  \multirow{6}{*}{} & \multirow{3}{*}{Yield} & in-plane & 16 & 6.3 & 50 & 18 \\ \cline{3-7}
  & & mid-plane & 9.3 & 3.9 & 37 & 13 \\ \cline{3-7}
  Background & & out-of-plane & 7.9 & 6.0 & 35 & 15  \\ \cline{2-7}
  fit & \multirow{3}{*}{Width} & in-plane & 23 & 4.2 & - & 12 \\ \cline{3-7}
  & & mid-plane & 25 & 2.7 & - & 23 \\ \cline{3-7}
  & & out-of-plane & 10 & 3.0 & - & 11 \\ \cline{3-7}
  \hline
\end{tabular}
\end{table}

\subsection{Impact of event plane resolution}\label{Sec:RxnPlaneResolutionEffects}

To understand the impact of a possible event plane dependence in the signal, we consider the Fourier decomposition approach to correcting for the event plane resolution as in~\cite{Adare:2018wjb}.  We can quantify the true azimuthal anistropy of the signal by a Fourier decomposition as

\begin{equation}\label{Eq:RxnPlaneResolutionCorrectionTrue}
 S(\Delta\phi) \big(1+2 \sum_{n = 1}^{\infty} v_{n}^{\mathrm{quench}} \cos (\phi^{\mathrm{trig}}-\psi)\big),
\end{equation}

\noindent where $S(\Delta\phi)$ is the correlation function of the signal averaged over all angles relative to the event plane, the $v_{n}^{\mathrm{quench}}$ are due to jet quenching, and $\phi^{\mathrm{trig}}$ is the azimuthal angle of the trigger particle.  The $v_{n}^{\mathrm{quench}}$ could also be a function of \dphi.  The measured azimuthal anisotropy of the signal is then given by

\begin{equation}\label{Eq:RxnPlaneResolutionCorrectionMeas}
 S \big(1+2 \sum_{n = 1}^{\infty} \Re_n v_{n}^{\mathrm{quench}} \cos (\phi^{\mathrm{trig}}-\psi)\big).
\end{equation}

\noindent Note that the $v_n^{\mathrm{quench}}$ are distinct from both the \vn from flow and the jet \vn.  The jet \vn measured in~\cite{Adam:2015mda,Aad:2013sla} are anisotropies in the number of jets relative to the event plane while the $v_n^{\mathrm{quench}}$ are a measure of the anisotropies of the constituents of those jets.  Precision extraction of the $v_{n}^{\mathrm{quench}}$ would require measurements of the signal in several bins of $\phi^{\mathrm{trig}}-\psi$ and is not feasible for this measurement.  \Eref{Eq:RxnPlaneResolutionCorrectionMeas} shows that the impact of the finite event plane resolution is small, since for this analysis $\Re_2 = 0.73$.

If we assume that the $v_{n}^{\mathrm{quench}}$ do not depend on \dphi, the yields are given by integrating the \eref{Eq:RxnPlaneResolutionCorrectionMeas} over the same angles of the trigger particle relative to the event plane.  The in-plane ($Y_{\mathrm {IP}}$), mid-plane ($Y_{\mathrm {MP}}$), and out-of-plane  ($Y_{\mathrm {OP}}$) yields up to $n = 3$ in terms of the average yield ($Y$) are then given by

\begin{align}
 Y_{\mathrm {IP}} & = Y (1+\frac{6}{\pi}\Re_1 v_1^{\mathrm{quench}} + \frac{3\sqrt{3}}{\pi}\Re_2 v_2^{\mathrm{quench}}+ \frac{4}{\pi}\Re_3 v_3^{\mathrm{quench}})\\ \nonumber 
 Y_{\mathrm {MP}} & = Y (1+\frac{6(\sqrt{3}-1)}{\pi}\Re_1 v_1^{\mathrm{quench}} - \frac{4}{\pi}\Re_3 v_3^{\mathrm{quench}})\\
 Y_{\mathrm {OP}} & = Y (1- \frac{3\sqrt{3}}{\pi}\Re_2 v_2^{\mathrm{quench}}).\nonumber
\end{align}

\noindent The differences and ratios of the yields up to $n = 3$ are then

\begin{align}\label{Eq:ApproxRatios}
 \frac{Y_{\mathrm {OP}}}{Y_{\mathrm {IP}}} & \approx  \frac{Y_{\mathrm {OP}} - Y_{\mathrm {IP}}}{Y} 
 \approx 1 - \frac{6}{\pi}\Re_1 v_1^{\mathrm{quench}}- \frac{6\sqrt{3}}{\pi}\Re_2 v_2^{\mathrm{quench}} - \frac{4}{\pi}\Re_3 v_3^{\mathrm{quench}}\\ \nonumber 
 \frac{Y_{\mathrm {MP}}}{Y_{\mathrm {IP}}} & \approx \frac{Y_{\mathrm {MP}} - Y_{\mathrm {IP}}}{Y} 
 \approx 1 - \frac{6\sqrt{3}-12}{\pi}\Re_1 v_1^{\mathrm{quench}} - \frac{3\sqrt{3}}{\pi}\Re_2 v_2^{\mathrm{quench}}- \frac{8}{\pi}\Re_3 v_3^{\mathrm{quench}}.  
\end{align}

\noindent Since the coefficients of the $v_n^{\mathrm{quench}}$ are on the order of one, the deviations of these ratios from one are on the order of the $v_n^{\mathrm{quench}}$.  The odd $v_n^{\mathrm{quench}}$ will partially cancel out because they will have opposite signs on different sides of the event plane and in the absence of surface bias, they will cancel out completely.  The $n = 2$ term is therefore likely the dominant term.  We use this expression to evaluate the approximate effect of the event plane resolution.  The allowed range of ${v}_{n}^{\mathrm {quench}}$ is $-0.5<{v}_{2}^{\mathrm {quench}}<0.5$, with positive (negative) values indicating suppression (enhancement) of the constituents.  While the ${v}_{n}^{\mathrm {quench}}$ is a measure of the asymmetry of modifications of jets relative to the event plane rather than the distribution of jets, we consider the asymmetry in the number of jets~\cite{Adam:2015mda,Aad:2013sla}, $v_{2}^{\mathrm {jet}} = 0.1$, as a reasonable value of ${v}_{n}^{\mathrm {quench}}$.  This would lead to $ \frac{Y_{\mathrm {OP}}}{Y_{\mathrm {IP}}} \approx \frac{Y_{\mathrm {OP}} - Y_{\mathrm {IP}}}{Y}  \approx 0.67$ and $ \frac{Y_{\mathrm {MP}}}{Y_{\mathrm {IP}}} \approx \frac{Y_{\mathrm {MP}} - Y_{\mathrm {IP}}}{Y} \approx 0.83$ with perfect event plane resolution ($\Re_{2} = 1.0$) and $ \frac{Y_{\mathrm {OP}}}{Y_{\mathrm {IP}}} \approx \frac{Y_{\mathrm {OP}} - Y_{\mathrm {IP}}}{Y} \approx 0.75$ and $ \frac{Y_{\mathrm {MP}}}{Y_{\mathrm {IP}}} \approx \frac{Y_{\mathrm {MP}} - Y_{\mathrm {IP}}}{Y} \approx 0.88$ with the event plane resolution in this analysis, $\Re_{2} = 0.73$.
 
\section{Results}\label{Sec:Results}

\begin{figure}[!htbp]
\begin{center}
\rotatebox{0}{\resizebox{\textwidth}{!}{
        \includegraphics{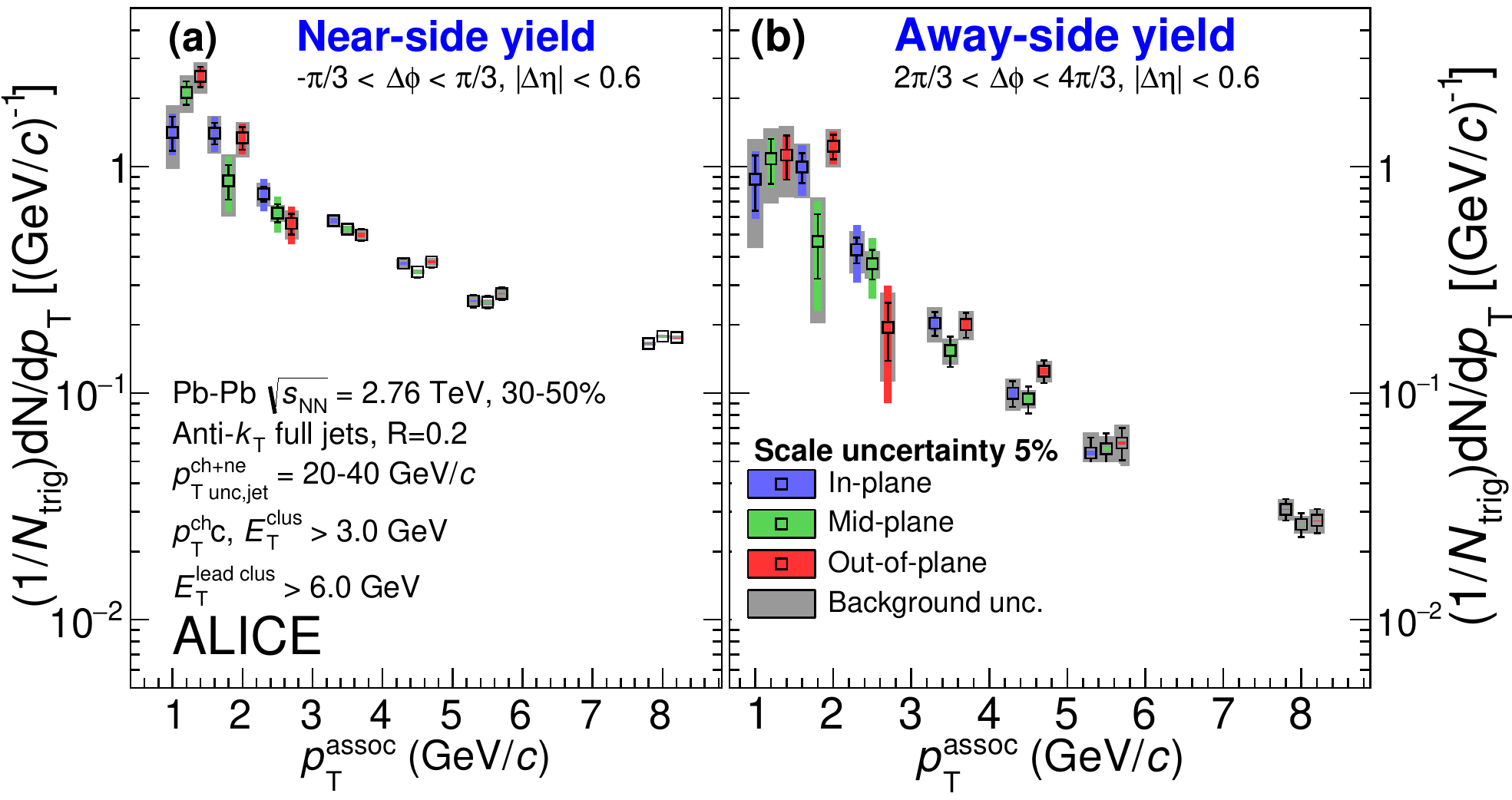}
}}
\caption{The (a) near-side and (b) away-side yield vs \ptassoc 
for \pttrigrange{20}{40} full jets of 30-50\% centrality in \PbPb collisions.  The wider band corresponds to the background uncertainty, which is non-trivially correlated point-to-point~\cite{Sharma:2015qra,Nattrass:2016cln}.  The narrower bands are the systematic uncertainties coming from the shape uncertainty of the acceptance correction.  There is an additional 5\% global scale uncertainty.  Points are displaced for visibility. 
}\label{fig:Yields_J20-40_C30-50}
\end{center}
\end{figure}

The near-side and away-side jet yields as a function of \ptassoc for \pttrigrange{20}{40} full jets in 30-50\% central \PbPb collisions are shown in~\fref{fig:Yields_J20-40_C30-50} for jets reconstructed in-plane, mid-plane, and out-of plane for \ptassocrange{1.0}{1.5}, \ptassocrange{1.5}{2.0}, \ptassocrange{2.0}{3.0}, \ptassocrange{3.0}{4.0}, \ptassocrange{4.0}{5.0}, \ptassocrange{5.0}{6.0}, and \ptassocrange{6.0}{10.0}.  The dominant feature is the decrease in the yield with increasing \ptassoc.  Note that yields with \ptassoc $> 3$ \GeV include jet constituents, complicating the interpretation of these data points.  We therefore focus on lower momentum on the near-side and on the away-side.

\begin{figure}[!htbp]
\begin{center}
\rotatebox{0}{\resizebox{\textwidth}{!}{
        \includegraphics{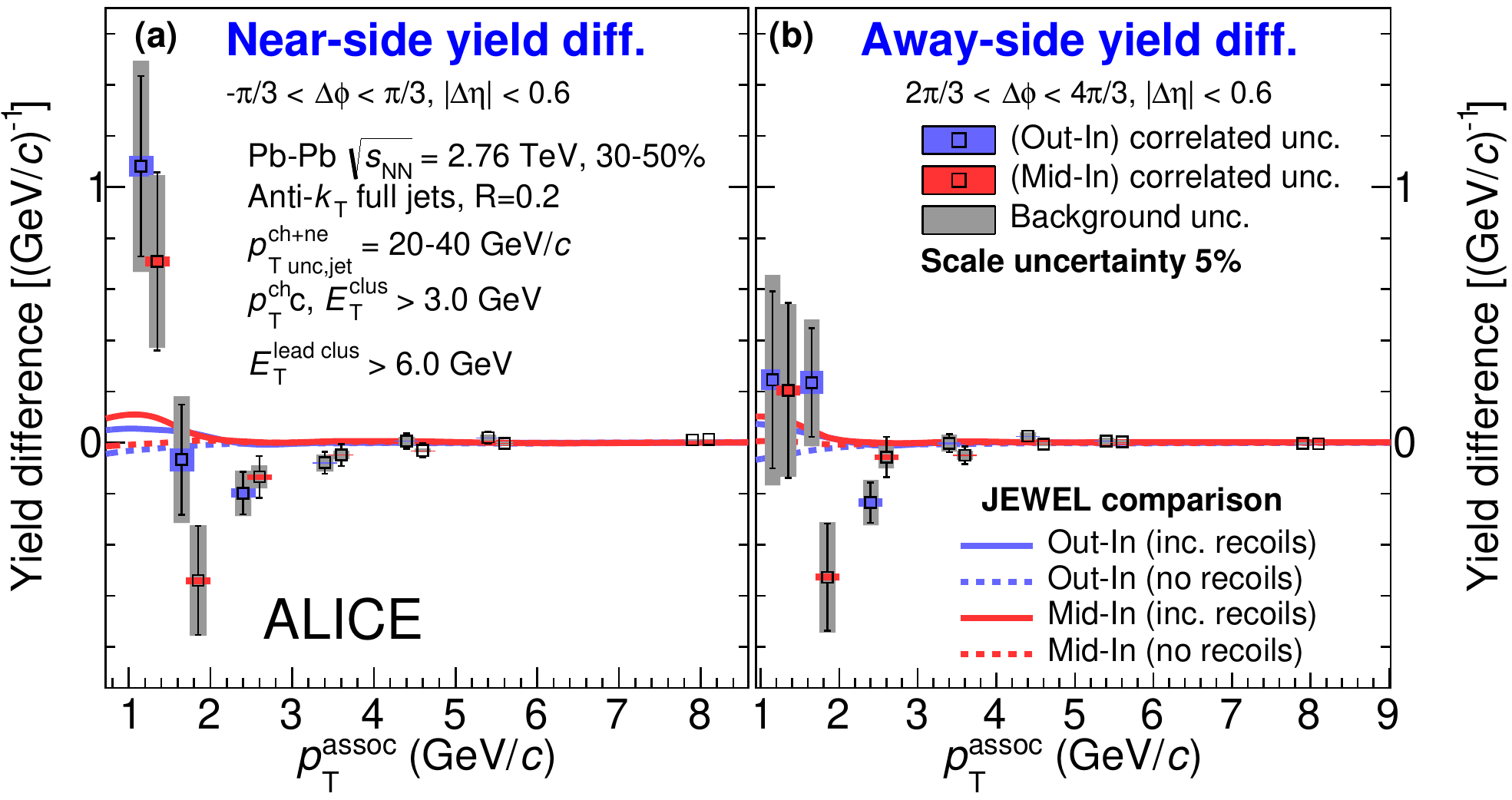}
}}
\caption{
The (a) near-side and (b) away-side yield differences vs \ptassoc 
for \pttrigrange{20}{40} full jets of 30-50\% centrality in \PbPb collisions.  The narrower band corresponds to the background uncertainty, which is non-trivially correlated point-to-point~\cite{Sharma:2015qra,Nattrass:2016cln}.  The wider bands are the systematic uncertainties coming from the shape uncertainty of the acceptance correction.  There is an additional 5\% global scale uncertainty.  Points are displaced for visibility.  Data are compared to calculations from JEWEL~\cite{Zapp:2013vla} with and without recoil particles.
}\label{fig:YieldDifferences_J20-40_C30-50}
\end{center}
\end{figure}

Jet-hadron correlations can be used to measure  changes in the momentum balance within the jet, as in~\cite{Adamczyk:2013jei}.  Partonic energy loss will shift energy in the jet from higher momentum constituents to lower momentum constituents, so if jets in-plane interact less with the medium, the differences $Y_{\mathrm{MP}} - Y_{\mathrm{IP}}$ and $Y_{\mathrm{OP}} - Y_{\mathrm{IP}}$ will be negative at high momenta and positive at low momenta.  For these differences, the systematic uncertainties partially cancel out. \Fref{fig:YieldDifferences_J20-40_C30-50} shows the yield differences $Y_{\mathrm{MP}} - Y_{\mathrm{IP}}$ and $Y_{\mathrm{OP}} - Y_{\mathrm{IP}}$ for the near- and \as.  There is no event plane dependence within uncertainties, consistent with expectations if $v_2^{\mathrm{quench}} \approx 0.1$ as observed for inclusive jet production.  Comparisons between yields in jet-hadron correlations in \AuAu and \pp collisions demonstrated suppression at high momenta and an enhancement at low momenta in \AuAu collisions at \sNN = 200~GeV~\cite{Adamczyk:2013jei}.  The lack of an event plane dependence therefore indicates that any dependence of these modifications on the average path length is less than our experimental uncertainties.

\begin{figure}[!htbp]
\begin{center}
\rotatebox{0}{\resizebox{\textwidth}{!}{
        \includegraphics{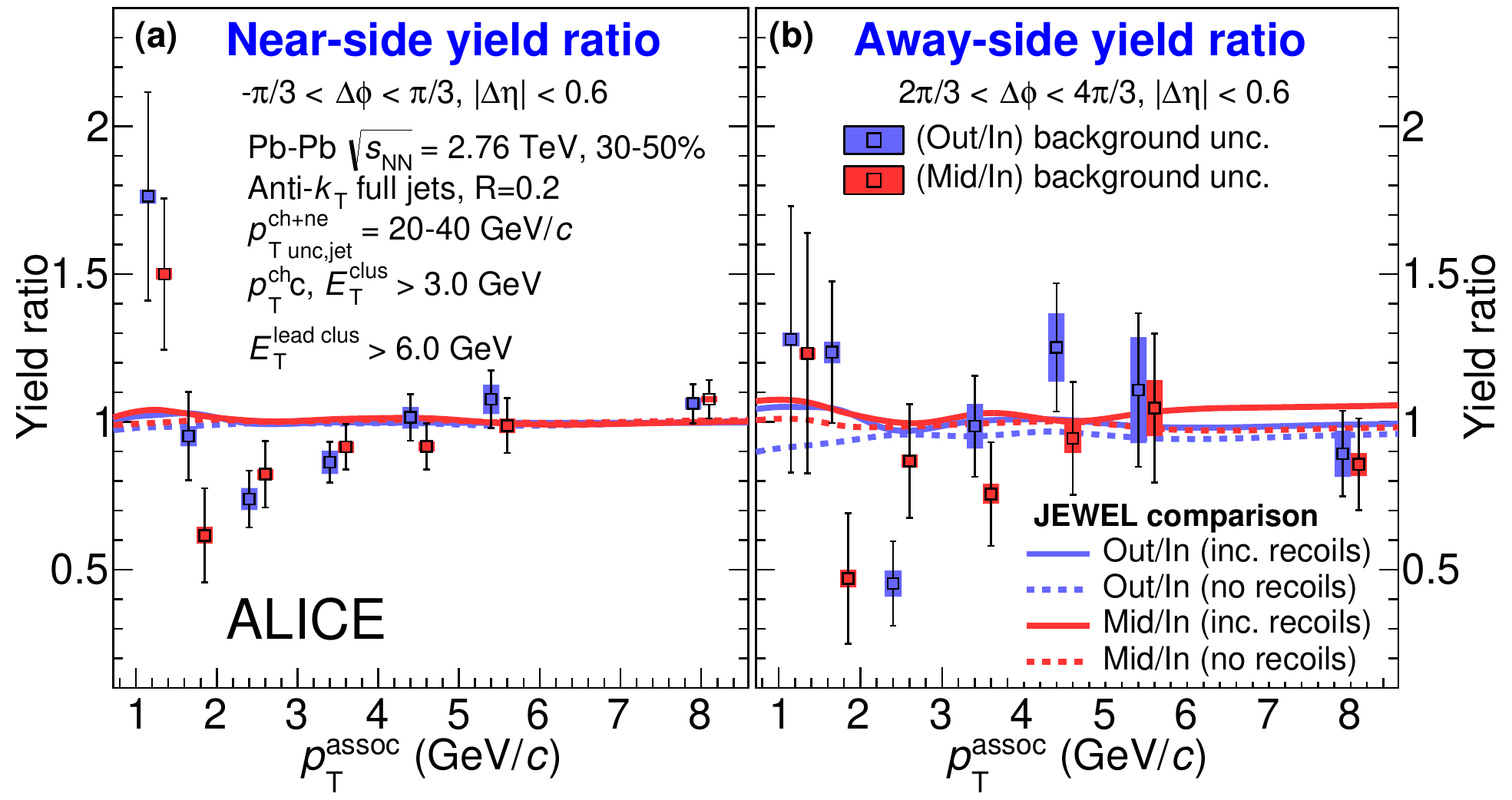}
}}
\caption{
The (a) near-side and (b) away-side yield ratios vs \ptassoc for \pttrigrange{20}{40} full jets of 30-50\% centrality in \PbPb collisions.  The bands correspond to the background uncertainty, which is non-trivially correlated point-to-point~\cite{Sharma:2015qra,Nattrass:2016cln}.  The systematic uncertainties coming from the shape uncertainty of the acceptance correction cancel out for the ratios.  Points are displaced for visibility.  Data are compared to calculations from JEWEL~\cite{Zapp:2013vla} with and without recoil particles.
}\label{fig:YieldRatios_J20-40_C30-50}
\end{center}
\end{figure}

To better quantify and examine the event plane dependence of the yields, ratios of mid-plane yields relative to in-plane yields $Y_{\mathrm{MP}} / Y_{\mathrm{IP}}$ and out-of-plane yields relative to in-plane yields $Y_{\mathrm{OP}} / Y_{\mathrm{IP}}$ as a function of \ptassoc are shown in~\fref{fig:YieldRatios_J20-40_C30-50} for both the near- and away-sides.  As for the yield differences, a substantial fraction of the systematic uncertainties cancel out for the ratios.  If medium modifications increase with increasing path length traversed by the parton, these ratios will be less than one at high momenta and greater than one at low momenta.  These ratios are consistent with one for all \ptassoc.  In contrast, $R_{\mathrm{AA}}$ can be as low as 0.1~\cite{Aamodt:2010jd}, indicating partonic energy loss.  

The modification of the correlated yield ratios $Y_{\mathrm{OP}} / Y_{\mathrm{IP}}$ and $Y_{\mathrm{MP}} / Y_{\mathrm{IP}}$ due to jet quenching can be estimated from \eref{Eq:ApproxRatios} as approximately $1 - 3.3 \times \Re_{2} v_{2}^{\mathrm {quench}}$ for out-of-plane to in-plane ratios and $1 - 1.7 \times \Re_{2} v_{2}^{\mathrm {quench}}$ for mid-plane to in-plane ratios following the logic in \sref{Sec:RxnPlaneResolutionEffects}.  
Since $\Re_{2} = 0.73$, the ratios in \fref{fig:YieldRatios_J20-40_C30-50} can be used to constrain a hypothetical $v_{2}^{\mathrm {quench}}$.  While $v_{2}^{\mathrm {quench}}$ is a measure of the azimuthal asymmetry in jet modifications rather than the number of jets, we use the asymmetry in the number of jets, $v_{2}^{\mathrm {jet}} = 0.1$~\cite{Adam:2015mda,Aad:2013sla}, as a reasonable value for $v_{2}^{\mathrm {quench}}$. If $v_{2}^{\mathrm {quench}}=0.1$, the out-of-plane to in-plane ratios would be 0.75 and the mid-plane to in-plane ratios would be 0.88.  The data in \fref{fig:YieldRatios_J20-40_C30-50} are therefore consistent both with $v_{2}^{\mathrm {quench}}$ comparable to the inclusive jet asymmetry and with no asymmetry.

To investigate whether or not there is a systematic change in the ratio of yields with the angle of the jet relative to the event plane, we fit the data in \fref{fig:YieldRatios_J20-40_C30-50} to a constant.  The systematic uncertainties are treated as uncorrelated point-to-point and added in quadrature to the statistical uncertainties.  The results are given in \tref{Tab:Fits} and are consistent with yield ratios of one.  We note that medium modifications could result in a \ptassoc dependence and this could be exacerbated by kinematic biases on the near side because associated particles with momenta above 3 \GeV are included in jet reconstruction.  
The $\chi^2$ per degree of freedom may be large either because this procedure averages over different physical effects which change with momentum or because of point-to-point correlations in the uncertainties.

\begin{table}
\caption{Results of fits to \fref{fig:YieldRatios_J20-40_C30-50} to a constant $c$, the $\chi^2$ over the number of degrees of freedom (NDF), the number of standard deviations $\sigma$ of $c$ from one, and the range of $c$ within a 90\% confidence limit (CL).}
\label{Tab:Fits}
\centering
\begin{tabular}{c | c c | c c }
\hline
          & \multicolumn{2}{c |}{Near-side}  & \multicolumn{2}{c}{Away-side}  \\
parameter & $Y_{\mathrm{OP}} / Y_{\mathrm{IP}}$  & $Y_{\mathrm{MP}} / Y_{\mathrm{IP}}$   & $Y_{\mathrm{OP}} / Y_{\mathrm{IP}}$  & $Y_{\mathrm{MP}} / Y_{\mathrm{IP}}$  \\ \hline
$c$ &0.972 $\pm$ 0.037 &0.960 $\pm$ 0.036 &0.885 $\pm$ 0.079 &0.835 $\pm$ 0.078   \\
$\chi^2$/NDF & 2.5  &2.4  &2.4  &0.8   \\ 
$\sigma$ & -0.8  &-1.1  &-1.5  &-2.1    \\ 
90\% CL &0.91 -- 1.03 &0.90 -- 1.02 &0.75 -- 1.02 &0.71 -- 0.96 \\
\hline
\end{tabular}
\end{table}

\begin{figure}[!htbp]
\begin{center}
\rotatebox{0}{\resizebox{\textwidth}{!}{
        \includegraphics{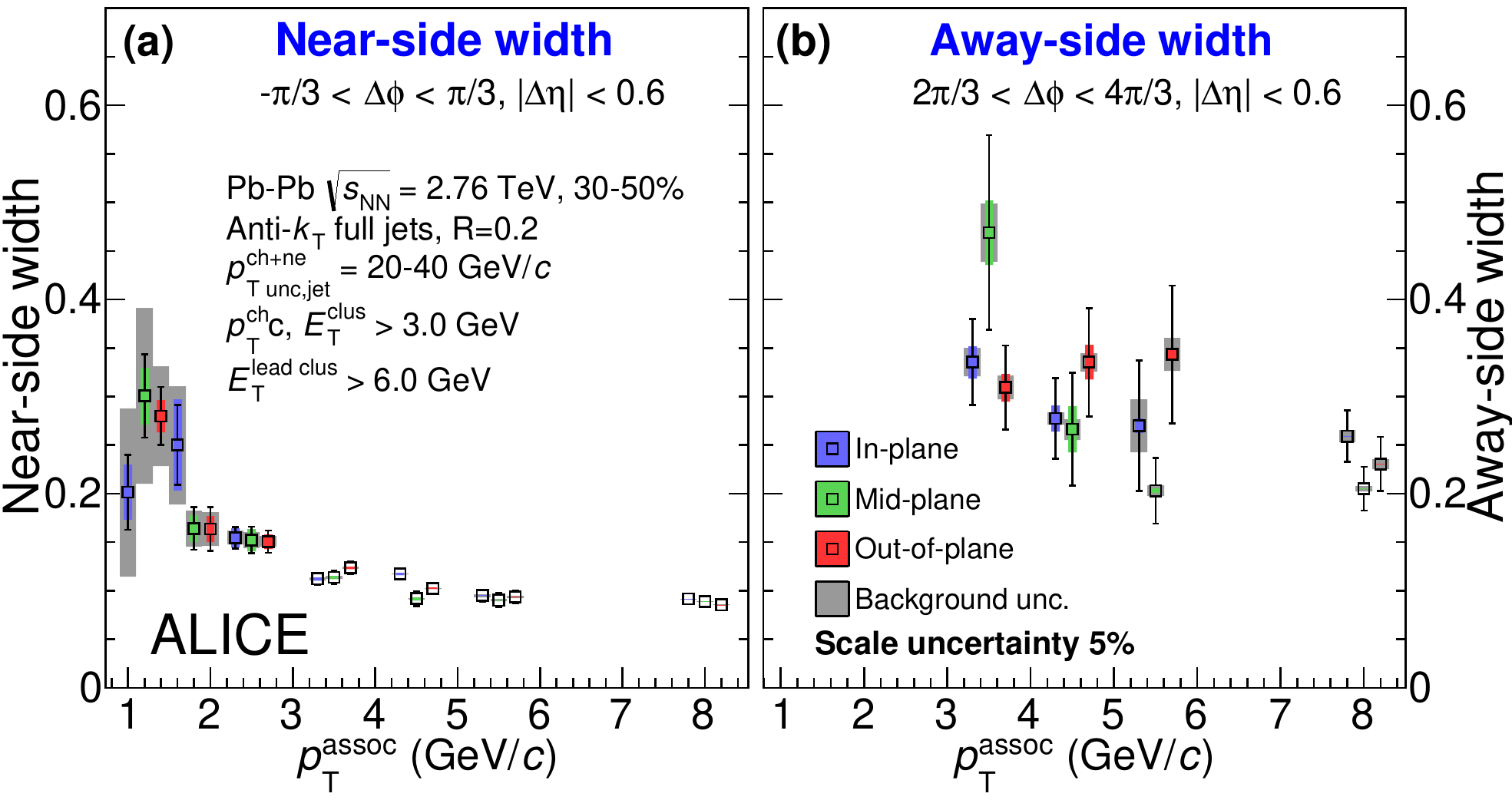}
}}
\caption{
The (a) near-side and (b) away-side widths vs \ptassoc for \pttrigrange{20}{40} full jets of 30-50\% centrality in \PbPb collisions.  The wider band corresponds the background uncertainty, which is non-trivially correlated point-to-point~\cite{Sharma:2015qra,Nattrass:2016cln}.  The narrower bands are the systematic uncertainties coming from the shape uncertainty of the acceptance correction.   Points are displaced for visibility.
}\label{fig:Widths_J20-40_C30-50}
\end{center}
\end{figure}

\Fref{fig:Widths_J20-40_C30-50} shows the widths from a fit to the Gaussian for the near- and \as.  Broadening would be expected from either collisional energy loss or gluon bremsstrahlung and path length dependent energy loss would lead to a greater width for jets out-of-plane than in-plane.  No event plane dependence is observed within uncertainties, indicating that any effect is smaller than the precision of the data.

The data in \fref{fig:YieldDifferences_J20-40_C30-50} and \fref{fig:YieldRatios_J20-40_C30-50} are compared to calculations from JEWEL, a jet energy loss model based on radiative and collisional energy loss in connection with partons sampled from a longitudinally expanding  medium~\cite{Zapp:2013vla}.  An important setting in the model is the choice of whether or not to keep the recoiled partons sampled from the medium in the simulation.  With no recoils, the lost jet momentum vanishes from the entire system, while including the recoils conserves the jet's overall momentum, but adds energy and background particles (from the medium) to the simulated dijet.  We compare to JEWEL with both recoils off and recoils on.  Results with recoils off are useful for modeling energy loss in the hard part of the jet.  Results with recoils on show where the jet's lost momentum goes.  Any experimental analysis would likely include some but not all of the recoil particles in JEWEL, as some proportion of the recoil particles are indistinguishable from background.
 
JEWEL only predicts a slight event plane dependence, despite the path length dependence of partonic energy loss, due to the dominant impact of jet-by-jet fluctuations in partonic energy loss over path length dependence~\cite{Milhano:2015mng,Zapp:2013zya}.  The slight event plane dependence predicted by JEWEL is well below the systematic uncertainty in the measurement.  The agreement of JEWEL with the data is therefore consistent with path length dependence having an insignificant impact compared to jet-by-jet fluctuations in energy loss, although fluctuations in the density of the medium (not included in the JEWEL model) may also suppress observable path length dependence.

\section{Conclusions}\label{Sec:Discussion}
Partonic interactions depend on the length traversed in the medium, so any medium modifications of the jet are expected to be path length dependent.  The path length traversed by a jet is correlated on average with the angle of the jet with respect to the event plane.  The use of the RPF method for background subtraction reduces the assumptions required for background subtraction and since the determination of the background is currently limited by statistics, it is likely that future studies could reduce these systematic uncertainties.  Measurements of jet-hadron correlations relative to the event plane in \PbPb collisions at \sNN = 2.76 TeV are presented.  Results are consistent with no dependence in the yields or the widths on the angle of the jet relative to the event plane within uncertainties.  
This may indicate that jet-by-jet fluctuations in partonic energy loss are important for a full description of medium modifications of jets.
 
\newenvironment{acknowledgement}{\relax}{\relax}
\begin{acknowledgement}
\section*{Acknowledgements}

The ALICE Collaboration would like to thank all its engineers and technicians for their invaluable contributions to the construction of the experiment and the CERN accelerator teams for the outstanding performance of the LHC complex.
The ALICE Collaboration gratefully acknowledges the resources and support provided by all Grid centres and the Worldwide LHC Computing Grid (WLCG) collaboration.
The ALICE Collaboration acknowledges the following funding agencies for their support in building and running the ALICE detector:
A. I. Alikhanyan National Science Laboratory (Yerevan Physics Institute) Foundation (ANSL), State Committee of Science and World Federation of Scientists (WFS), Armenia;
Austrian Academy of Sciences, Austrian Science Fund (FWF): [M 2467-N36] and Nationalstiftung f\"{u}r Forschung, Technologie und Entwicklung, Austria;
Ministry of Communications and High Technologies, National Nuclear Research Center, Azerbaijan;
Conselho Nacional de Desenvolvimento Cient\'{\i}fico e Tecnol\'{o}gico (CNPq), Financiadora de Estudos e Projetos (Finep), Funda\c{c}\~{a}o de Amparo \`{a} Pesquisa do Estado de S\~{a}o Paulo (FAPESP) and Universidade Federal do Rio Grande do Sul (UFRGS), Brazil;
Ministry of Education of China (MOEC) , Ministry of Science \& Technology of China (MSTC) and National Natural Science Foundation of China (NSFC), China;
Ministry of Science and Education and Croatian Science Foundation, Croatia;
Centro de Aplicaciones Tecnol\'{o}gicas y Desarrollo Nuclear (CEADEN), Cubaenerg\'{\i}a, Cuba;
Ministry of Education, Youth and Sports of the Czech Republic, Czech Republic;
The Danish Council for Independent Research | Natural Sciences, the VILLUM FONDEN and Danish National Research Foundation (DNRF), Denmark;
Helsinki Institute of Physics (HIP), Finland;
Commissariat \`{a} l'Energie Atomique (CEA), Institut National de Physique Nucl\'{e}aire et de Physique des Particules (IN2P3) and Centre National de la Recherche Scientifique (CNRS) and R\'{e}gion des  Pays de la Loire, France;
Bundesministerium f\"{u}r Bildung und Forschung (BMBF) and GSI Helmholtzzentrum f\"{u}r Schwerionenforschung GmbH, Germany;
General Secretariat for Research and Technology, Ministry of Education, Research and Religions, Greece;
National Research, Development and Innovation Office, Hungary;
Department of Atomic Energy Government of India (DAE), Department of Science and Technology, Government of India (DST), University Grants Commission, Government of India (UGC) and Council of Scientific and Industrial Research (CSIR), India;
Indonesian Institute of Science, Indonesia;
Centro Fermi - Museo Storico della Fisica e Centro Studi e Ricerche Enrico Fermi and Istituto Nazionale di Fisica Nucleare (INFN), Italy;
Institute for Innovative Science and Technology , Nagasaki Institute of Applied Science (IIST), Japanese Ministry of Education, Culture, Sports, Science and Technology (MEXT) and Japan Society for the Promotion of Science (JSPS) KAKENHI, Japan;
Consejo Nacional de Ciencia (CONACYT) y Tecnolog\'{i}a, through Fondo de Cooperaci\'{o}n Internacional en Ciencia y Tecnolog\'{i}a (FONCICYT) and Direcci\'{o}n General de Asuntos del Personal Academico (DGAPA), Mexico;
Nederlandse Organisatie voor Wetenschappelijk Onderzoek (NWO), Netherlands;
The Research Council of Norway, Norway;
Commission on Science and Technology for Sustainable Development in the South (COMSATS), Pakistan;
Pontificia Universidad Cat\'{o}lica del Per\'{u}, Peru;
Ministry of Science and Higher Education and National Science Centre, Poland;
Korea Institute of Science and Technology Information and National Research Foundation of Korea (NRF), Republic of Korea;
Ministry of Education and Scientific Research, Institute of Atomic Physics and Ministry of Research and Innovation and Institute of Atomic Physics, Romania;
Joint Institute for Nuclear Research (JINR), Ministry of Education and Science of the Russian Federation, National Research Centre Kurchatov Institute, Russian Science Foundation and Russian Foundation for Basic Research, Russia;
Ministry of Education, Science, Research and Sport of the Slovak Republic, Slovakia;
National Research Foundation of South Africa, South Africa;
Swedish Research Council (VR) and Knut \& Alice Wallenberg Foundation (KAW), Sweden;
European Organization for Nuclear Research, Switzerland;
Suranaree University of Technology (SUT), National Science and Technology Development Agency (NSDTA) and Office of the Higher Education Commission under NRU project of Thailand, Thailand;
Turkish Atomic Energy Agency (TAEK), Turkey;
National Academy of  Sciences of Ukraine, Ukraine;
Science and Technology Facilities Council (STFC), United Kingdom;
National Science Foundation of the United States of America (NSF) and United States Department of Energy, Office of Nuclear Physics (DOE NP), United States of America.   
\end{acknowledgement}

\bibliographystyle{utphys} 
\bibliography{references/Bibliography.bib,references/references.bib}

\newpage
\appendix

\section{The ALICE Collaboration}
\label{app:collab}

\begingroup
\small
\begin{flushleft}
S.~Acharya\Irefn{org141}\And 
D.~Adamov\'{a}\Irefn{org93}\And 
A.~Adler\Irefn{org73}\And 
J.~Adolfsson\Irefn{org79}\And 
M.M.~Aggarwal\Irefn{org98}\And 
G.~Aglieri Rinella\Irefn{org34}\And 
M.~Agnello\Irefn{org31}\And 
N.~Agrawal\Irefn{org10}\textsuperscript{,}\Irefn{org53}\And 
Z.~Ahammed\Irefn{org141}\And 
S.~Ahmad\Irefn{org17}
% ...
\And
\renewcommand\labelenumi{\textsuperscript{\theenumi}~}

\section*{Affiliation notes}
\renewcommand\theenumi{\roman{enumi}}
\begin{Authlist}
\item \Adef{org*}Deceased
\item \Adef{orgI}Dipartimento DET del Politecnico di Torino, Turin, Italy
\item \Adef{orgII}M.V. Lomonosov Moscow State University, D.V. Skobeltsyn Institute of Nuclear, Physics, Moscow, Russia
\item \Adef{orgIII}Department of Applied Physics, Aligarh Muslim University, Aligarh, India
\item \Adef{orgIV}Institute of Theoretical Physics, University of Wroclaw, Poland
\end{Authlist}

\section*{Collaboration Institutes}
\renewcommand\theenumi{\arabic{enumi}~}
\begin{Authlist}
\item \Idef{org1}A.I. Alikhanyan National Science Laboratory (Yerevan Physics Institute) Foundation, Yerevan, Armenia
\item \Idef{org2}Bogolyubov Institute for Theoretical Physics, National Academy of Sciences of Ukraine, Kiev, Ukraine
\item \Idef{org3}Bose Institute, Department of Physics  and Centre for Astroparticle Physics and Space Science (CAPSS), Kolkata, India
\item \Idef{org4}Budker Institute for Nuclear Physics, Novosibirsk, Russia
\item \Idef{org5}California Polytechnic State University, San Luis Obispo, California, United States
\item \Idef{org6}Central China Normal University, Wuhan, China
\item \Idef{org7}Centre de Calcul de l'IN2P3, Villeurbanne, Lyon, France
\item \Idef{org8}Centro de Aplicaciones Tecnol\'{o}gicas y Desarrollo Nuclear (CEADEN), Havana, Cuba
\item \Idef{org9}Centro de Investigaci\'{o}n y de Estudios Avanzados (CINVESTAV), Mexico City and M\'{e}rida, Mexico
\item \Idef{org10}Centro Fermi - Museo Storico della Fisica e Centro Studi e Ricerche ``Enrico Fermi', Rome, Italy
\item \Idef{org11}Chicago State University, Chicago, Illinois, United States
\item \Idef{org12}China Institute of Atomic Energy, Beijing, China
\item \Idef{org13}Chonbuk National University, Jeonju, Republic of Korea
\item \Idef{org14}Comenius University Bratislava, Faculty of Mathematics, Physics and Informatics, Bratislava, Slovakia
\item \Idef{org15}COMSATS University Islamabad, Islamabad, Pakistan
\item \Idef{org16}Creighton University, Omaha, Nebraska, United States
\item \Idef{org17}Department of Physics, Aligarh Muslim University, Aligarh, India
\item \Idef{org18}Department of Physics, Pusan National University, Pusan, Republic of Korea
\item \Idef{org19}Department of Physics, Sejong University, Seoul, Republic of Korea
\item \Idef{org20}Department of Physics, University of California, Berkeley, California, United States
\item \Idef{org21}Department of Physics, University of Oslo, Oslo, Norway
\item \Idef{org22}Department of Physics and Technology, University of Bergen, Bergen, Norway
\item \Idef{org23}Dipartimento di Fisica dell'Universit\`{a} 'La Sapienza' and Sezione INFN, Rome, Italy
\item \Idef{org24}Dipartimento di Fisica dell'Universit\`{a} and Sezione INFN, Cagliari, Italy
\item \Idef{org25}Dipartimento di Fisica dell'Universit\`{a} and Sezione INFN, Trieste, Italy
\item \Idef{org26}Dipartimento di Fisica dell'Universit\`{a} and Sezione INFN, Turin, Italy
\item \Idef{org27}Dipartimento di Fisica e Astronomia dell'Universit\`{a} and Sezione INFN, Bologna, Italy
\item \Idef{org28}Dipartimento di Fisica e Astronomia dell'Universit\`{a} and Sezione INFN, Catania, Italy
\item \Idef{org29}Dipartimento di Fisica e Astronomia dell'Universit\`{a} and Sezione INFN, Padova, Italy
\item \Idef{org30}Dipartimento di Fisica `E.R.~Caianiello' dell'Universit\`{a} and Gruppo Collegato INFN, Salerno, Italy
\item \Idef{org31}Dipartimento DISAT del Politecnico and Sezione INFN, Turin, Italy
\item \Idef{org32}Dipartimento di Scienze e Innovazione Tecnologica dell'Universit\`{a} del Piemonte Orientale and INFN Sezione di Torino, Alessandria, Italy
\item \Idef{org33}Dipartimento Interateneo di Fisica `M.~Merlin' and Sezione INFN, Bari, Italy
\item \Idef{org34}European Organization for Nuclear Research (CERN), Geneva, Switzerland
\item \Idef{org35}Faculty of Electrical Engineering, Mechanical Engineering and Naval Architecture, University of Split, Split, Croatia
\item \Idef{org36}Faculty of Engineering and Science, Western Norway University of Applied Sciences, Bergen, Norway
\item \Idef{org37}Faculty of Nuclear Sciences and Physical Engineering, Czech Technical University in Prague, Prague, Czech Republic
\item \Idef{org38}Faculty of Science, P.J.~\v{S}af\'{a}rik University, Ko\v{s}ice, Slovakia
\item \Idef{org39}Frankfurt Institute for Advanced Studies, Johann Wolfgang Goethe-Universit\"{a}t Frankfurt, Frankfurt, Germany
\item \Idef{org40}Gangneung-Wonju National University, Gangneung, Republic of Korea
\item \Idef{org41}Gauhati University, Department of Physics, Guwahati, India
\item \Idef{org42}Helmholtz-Institut f\"{u}r Strahlen- und Kernphysik, Rheinische Friedrich-Wilhelms-Universit\"{a}t Bonn, Bonn, Germany
\item \Idef{org43}Helsinki Institute of Physics (HIP), Helsinki, Finland
\item \Idef{org44}High Energy Physics Group,  Universidad Aut\'{o}noma de Puebla, Puebla, Mexico
\item \Idef{org45}Hiroshima University, Hiroshima, Japan
\item \Idef{org46}Hochschule Worms, Zentrum  f\"{u}r Technologietransfer und Telekommunikation (ZTT), Worms, Germany
\item \Idef{org47}Horia Hulubei National Institute of Physics and Nuclear Engineering, Bucharest, Romania
\item \Idef{org48}Indian Institute of Technology Bombay (IIT), Mumbai, India
\item \Idef{org49}Indian Institute of Technology Indore, Indore, India
\item \Idef{org50}Indonesian Institute of Sciences, Jakarta, Indonesia
\item \Idef{org51}INFN, Laboratori Nazionali di Frascati, Frascati, Italy
\item \Idef{org52}INFN, Sezione di Bari, Bari, Italy
\item \Idef{org53}INFN, Sezione di Bologna, Bologna, Italy
\item \Idef{org54}INFN, Sezione di Cagliari, Cagliari, Italy
\item \Idef{org55}INFN, Sezione di Catania, Catania, Italy
\item \Idef{org56}INFN, Sezione di Padova, Padova, Italy
\item \Idef{org57}INFN, Sezione di Roma, Rome, Italy
\item \Idef{org58}INFN, Sezione di Torino, Turin, Italy
\item \Idef{org59}INFN, Sezione di Trieste, Trieste, Italy
\item \Idef{org60}Inha University, Incheon, Republic of Korea
\item \Idef{org61}Institut de Physique Nucl\'{e}aire d'Orsay (IPNO), Institut National de Physique Nucl\'{e}aire et de Physique des Particules (IN2P3/CNRS), Universit\'{e} de Paris-Sud, Universit\'{e} Paris-Saclay, Orsay, France
\item \Idef{org62}Institute for Nuclear Research, Academy of Sciences, Moscow, Russia
\item \Idef{org63}Institute for Subatomic Physics, Utrecht University/Nikhef, Utrecht, Netherlands
\item \Idef{org64}Institute of Experimental Physics, Slovak Academy of Sciences, Ko\v{s}ice, Slovakia
\item \Idef{org65}Institute of Physics, Homi Bhabha National Institute, Bhubaneswar, India
\item \Idef{org66}Institute of Physics of the Czech Academy of Sciences, Prague, Czech Republic
\item \Idef{org67}Institute of Space Science (ISS), Bucharest, Romania
\item \Idef{org68}Institut f\"{u}r Kernphysik, Johann Wolfgang Goethe-Universit\"{a}t Frankfurt, Frankfurt, Germany
\item \Idef{org69}Instituto de Ciencias Nucleares, Universidad Nacional Aut\'{o}noma de M\'{e}xico, Mexico City, Mexico
\item \Idef{org70}Instituto de F\'{i}sica, Universidade Federal do Rio Grande do Sul (UFRGS), Porto Alegre, Brazil
\item \Idef{org71}Instituto de F\'{\i}sica, Universidad Nacional Aut\'{o}noma de M\'{e}xico, Mexico City, Mexico
\item \Idef{org72}iThemba LABS, National Research Foundation, Somerset West, South Africa
\item \Idef{org73}Johann-Wolfgang-Goethe Universit\"{a}t Frankfurt Institut f\"{u}r Informatik, Fachbereich Informatik und Mathematik, Frankfurt, Germany
\item \Idef{org74}Joint Institute for Nuclear Research (JINR), Dubna, Russia
\item \Idef{org75}Korea Institute of Science and Technology Information, Daejeon, Republic of Korea
\item \Idef{org76}KTO Karatay University, Konya, Turkey
\item \Idef{org77}Laboratoire de Physique Subatomique et de Cosmologie, Universit\'{e} Grenoble-Alpes, CNRS-IN2P3, Grenoble, France
\item \Idef{org78}Lawrence Berkeley National Laboratory, Berkeley, California, United States
\item \Idef{org79}Lund University Department of Physics, Division of Particle Physics, Lund, Sweden
\item \Idef{org80}Nagasaki Institute of Applied Science, Nagasaki, Japan
\item \Idef{org81}Nara Women{'}s University (NWU), Nara, Japan
\item \Idef{org82}National and Kapodistrian University of Athens, School of Science, Department of Physics , Athens, Greece
\item \Idef{org83}National Centre for Nuclear Research, Warsaw, Poland
\item \Idef{org84}National Institute of Science Education and Research, Homi Bhabha National Institute, Jatni, India
\item \Idef{org85}National Nuclear Research Center, Baku, Azerbaijan
\item \Idef{org86}National Research Centre Kurchatov Institute, Moscow, Russia
\item \Idef{org87}Niels Bohr Institute, University of Copenhagen, Copenhagen, Denmark
\item \Idef{org88}Nikhef, National institute for subatomic physics, Amsterdam, Netherlands
\item \Idef{org89}NRC Kurchatov Institute IHEP, Protvino, Russia
\item \Idef{org90}NRC «Kurchatov Institute»  - ITEP, Moscow, Russia
\item \Idef{org91}NRNU Moscow Engineering Physics Institute, Moscow, Russia
\item \Idef{org92}Nuclear Physics Group, STFC Daresbury Laboratory, Daresbury, United Kingdom
\item \Idef{org93}Nuclear Physics Institute of the Czech Academy of Sciences, \v{R}e\v{z} u Prahy, Czech Republic
\item \Idef{org94}Oak Ridge National Laboratory, Oak Ridge, Tennessee, United States
\item \Idef{org95}Ohio State University, Columbus, Ohio, United States
\item \Idef{org96}Petersburg Nuclear Physics Institute, Gatchina, Russia
\item \Idef{org97}Physics department, Faculty of science, University of Zagreb, Zagreb, Croatia
\item \Idef{org98}Physics Department, Panjab University, Chandigarh, India
\item \Idef{org99}Physics Department, University of Jammu, Jammu, India
\item \Idef{org100}Physics Department, University of Rajasthan, Jaipur, India
\item \Idef{org101}Physikalisches Institut, Eberhard-Karls-Universit\"{a}t T\"{u}bingen, T\"{u}bingen, Germany
\item \Idef{org102}Physikalisches Institut, Ruprecht-Karls-Universit\"{a}t Heidelberg, Heidelberg, Germany
\item \Idef{org103}Physik Department, Technische Universit\"{a}t M\"{u}nchen, Munich, Germany
\item \Idef{org104}Politecnico di Bari, Bari, Italy
\item \Idef{org105}Research Division and ExtreMe Matter Institute EMMI, GSI Helmholtzzentrum f\"ur Schwerionenforschung GmbH, Darmstadt, Germany
\item \Idef{org106}Rudjer Bo\v{s}kovi\'{c} Institute, Zagreb, Croatia
\item \Idef{org107}Russian Federal Nuclear Center (VNIIEF), Sarov, Russia
\item \Idef{org108}Saha Institute of Nuclear Physics, Homi Bhabha National Institute, Kolkata, India
\item \Idef{org109}School of Physics and Astronomy, University of Birmingham, Birmingham, United Kingdom
\item \Idef{org110}Secci\'{o}n F\'{\i}sica, Departamento de Ciencias, Pontificia Universidad Cat\'{o}lica del Per\'{u}, Lima, Peru
\item \Idef{org111}Shanghai Institute of Applied Physics, Shanghai, China
\item \Idef{org112}St. Petersburg State University, St. Petersburg, Russia
\item \Idef{org113}Stefan Meyer Institut f\"{u}r Subatomare Physik (SMI), Vienna, Austria
\item \Idef{org114}SUBATECH, IMT Atlantique, Universit\'{e} de Nantes, CNRS-IN2P3, Nantes, France
\item \Idef{org115}Suranaree University of Technology, Nakhon Ratchasima, Thailand
\item \Idef{org116}Technical University of Ko\v{s}ice, Ko\v{s}ice, Slovakia
\item \Idef{org117}Technische Universit\"{a}t M\"{u}nchen, Excellence Cluster 'Universe', Munich, Germany
\item \Idef{org118}The Henryk Niewodniczanski Institute of Nuclear Physics, Polish Academy of Sciences, Cracow, Poland
\item \Idef{org119}The University of Texas at Austin, Austin, Texas, United States
\item \Idef{org120}Universidad Aut\'{o}noma de Sinaloa, Culiac\'{a}n, Mexico
\item \Idef{org121}Universidade de S\~{a}o Paulo (USP), S\~{a}o Paulo, Brazil
\item \Idef{org122}Universidade Estadual de Campinas (UNICAMP), Campinas, Brazil
\item \Idef{org123}Universidade Federal do ABC, Santo Andre, Brazil
\item \Idef{org124}University of Cape Town, Cape Town, South Africa
\item \Idef{org125}University of Houston, Houston, Texas, United States
\item \Idef{org126}University of Jyv\"{a}skyl\"{a}, Jyv\"{a}skyl\"{a}, Finland
\item \Idef{org127}University of Liverpool, Liverpool, United Kingdom
\item \Idef{org128}University of Science and Techonology of China, Hefei, China
\item \Idef{org129}University of South-Eastern Norway, Tonsberg, Norway
\item \Idef{org130}University of Tennessee, Knoxville, Tennessee, United States
\item \Idef{org131}University of the Witwatersrand, Johannesburg, South Africa
\item \Idef{org132}University of Tokyo, Tokyo, Japan
\item \Idef{org133}University of Tsukuba, Tsukuba, Japan
\item \Idef{org134}Universit\'{e} Clermont Auvergne, CNRS/IN2P3, LPC, Clermont-Ferrand, France
\item \Idef{org135}Universit\'{e} de Lyon, Universit\'{e} Lyon 1, CNRS/IN2P3, IPN-Lyon, Villeurbanne, Lyon, France
\item \Idef{org136}Universit\'{e} de Strasbourg, CNRS, IPHC UMR 7178, F-67000 Strasbourg, France, Strasbourg, France
\item \Idef{org137}Universit\'{e} Paris-Saclay Centre d'Etudes de Saclay (CEA), IRFU, D\'{e}partment de Physique Nucl\'{e}aire (DPhN), Saclay, France
\item \Idef{org138}Universit\`{a} degli Studi di Foggia, Foggia, Italy
\item \Idef{org139}Universit\`{a} degli Studi di Pavia, Pavia, Italy
\item \Idef{org140}Universit\`{a} di Brescia, Brescia, Italy
\item \Idef{org141}Variable Energy Cyclotron Centre, Homi Bhabha National Institute, Kolkata, India
\item \Idef{org142}Warsaw University of Technology, Warsaw, Poland
\item \Idef{org143}Wayne State University, Detroit, Michigan, United States
\item \Idef{org144}Westf\"{a}lische Wilhelms-Universit\"{a}t M\"{u}nster, Institut f\"{u}r Kernphysik, M\"{u}nster, Germany
\item \Idef{org145}Wigner Research Centre for Physics, Budapest, Hungary
\item \Idef{org146}Yale University, New Haven, Connecticut, United States
\item \Idef{org147}Yonsei University, Seoul, Republic of Korea
\end{Authlist}
\endgroup
\end{document}